\documentclass[sigconf]{acmart}

\copyrightyear{2023}
\acmYear{2023}
\setcopyright{acmlicensed}\acmConference[RAID '23]{The 26th International Symposium on Research in Attacks, Intrusions and Defenses}{October 16--18, 2023}{Hong Kong, Hong Kong}
\acmBooktitle{The 26th International Symposium on Research in Attacks, Intrusions and Defenses (RAID '23), October 16--18, 2023, Hong Kong, Hong Kong}
\acmPrice{15.00}
\acmDOI{10.1145/3607199.3607226}
\acmISBN{979-8-4007-0765-0/23/10}

\usepackage{pifont}% http://ctan.org/pkg/pifont
\newcommand{\cmark}{\ding{51}}%
\newcommand{\xmark}{\ding{55}}%

\usepackage{listings}
\definecolor{codegreen}{rgb}{0,0.6,0}
\definecolor{codegray}{rgb}{0.5,0.5,0.5}
\definecolor{codepurple}{rgb}{0.58,0,0.82}
\definecolor{backcolour}{rgb}{0.95,0.95,0.92}
\lstdefinestyle{mystyle}{
    commentstyle=\color{codegreen},
    keywordstyle=\color{magenta},
    numberstyle=\tiny\color{codegray},
    stringstyle=\color{codepurple},
    basicstyle=\ttfamily\footnotesize,
    breakatwhitespace=false,         
    breaklines=true,                 
    captionpos=b,                    
    keepspaces=true,                 
    numbers=left,                    
    numbersep=5pt,                  
    showspaces=false,                
    showstringspaces=false,
    showtabs=false,                  
    tabsize=2,
    columns=fullflexible
}
\usepackage{caption}

\usepackage{array,multirow,graphicx}

\usepackage{tabularx}
\usepackage{booktabs}
\usepackage{wasysym}

\usepackage{enumitem}
\usepackage{tikz}

\usepackage{hyperref}

\definecolor{blockColor}{RGB}{0,0,0}

\definecolor{datagramColor}{RGB}{64,64,64}

\definecolor{channelColor}{RGB}{127,127,127}

\definecolor{serviceColor}{RGB}{166,166,166}

\definecolor{tagsColor}{RGB}{217,217,217}

\usepackage{soul}

\begin{document}

% USENIX: don't want date printed
\date{}

% Framework name
\newcommand{\FRAMEWORKNAME}{FieldFuzz}
\newcommand{\CVECOUNT}{three}

%make title bold and 14 pt font (Latex default is non-bold, 16 pt)
% \title{\Large \bf FieldFuzz: Towards vulnerability discovery in ICS field devices by leveraging native components of the control runtime}
% \title{\Large \bf FieldFuzz: Enabling vulnerability discovery in Industrial Control Systems supply chain using stateful system-level fuzzing}
% \title{FieldFuzz: Stateful Fuzzing of Proprietary Embedded Systems using Injected Monitors }
% \title{FieldFuzz: Stateful Fuzzing of Proprietary Embedded Systems using Injected Ghosts }
% \title{FieldFuzz: Stateful Fuzzing of Proprietary Industrial Controllers using Injected Ghosts}
% \title{FieldFuzz: Deep In Situ Fuzzing of Proprietary Industrial Controllers via the Network}
\title{FieldFuzz: In Situ Blackbox Fuzzing of Proprietary Industrial Automation Runtimes via the Network}
%\title{Ghost In The Shell: Exploring Control Logic Vulnerabilities on Industrial Controllers}

% Page limit: 13 typeset pages, excluding bibliography and appendices

% for over three affiliations, or if they all won't fit within the width
% of the page, use this alternative format:
% 
% \author{
% \IEEEauthorblockN{
% Andrei Bytes\IEEEauthorrefmark{1},
% Prashant Hari Narayan Rajput\IEEEauthorrefmark{2},
% Constantine Doumanidis\IEEEauthorrefmark{3}, \\
% Michail Maniatakos\IEEEauthorrefmark{3},
% Jianying Zhou\IEEEauthorrefmark{1} and
% Nils Ole Tippenhauer\IEEEauthorrefmark{4}}

% \IEEEauthorblockA{\IEEEauthorrefmark{1}Singapore University of Technology and Design}

% \IEEEauthorblockA{\IEEEauthorrefmark{2}NYU Tandon School of Engineering}

% \IEEEauthorblockA{\IEEEauthorrefmark{3}New York University Abu Dhabi}

% \IEEEauthorblockA{\IEEEauthorrefmark{4}CISPA Helmholtz Center for Information Security}}

% \author{\IEEEauthorblockN{Anonymous}}

% \affil[ ]{\textit {andrei\_bytes@mymail.sutd.edu.sg, prashanthrajput@nyu.edu, michail.maniatakos@nyu.edu, jianying_zhou@sutd.edu.sg}}

% Andrei Bytes (Singapore University of Technology and Design) <andrei_bytes@mymail.sutd.edu.sg>

% Prashant Hari Narayan Rajput (NYU Tandon School of Engineering) <prashanthrajput@nyu.edu>

% Michail Maniatakos (New York University Abu Dhabi) <michail.maniatakos@nyu.edu>

% Jianying Zhou (Singapore University of Technology and Design) <jianying_zhou@sutd.edu.sg>

\author{Andrei Bytes}
% \email{andrei\_bytes@mymail.sutd.edu.sg}
\affiliation{%
  \institution{Singapore University of Technology and Design}
  \country{Singapore}
}

\author{Prashant Hari Narayan Rajput}
\affiliation{%
  \institution{Tandon School of Engineering}
  \city{Brooklyn}
  \state{New York}
  \country{USA}
}

\author{Constantine Doumanidis}
\affiliation{%
  \institution{New York University Abu Dhabi}
  \city{Abu Dhabi}
  \country{UAE}
}

\author{Nils Ole Tippenhauer}
\affiliation{%
  \institution{CISPA Helmholtz Center for Information Security}
  \country{Saarbr\"ucken, Germany}
}

\author{Michail Maniatakos}
\affiliation{%
  \institution{New York University Abu Dhabi}
  \city{Abu Dhabi}
  \country{UAE}
}

\author{Jianying Zhou}
\affiliation{%
  \institution{Singapore University of Technology and Design}
  \country{Singapore}
}

\renewcommand{\shortauthors}{Bytes et al.}

\acmConference[RAID '23]{ACM Conference}{October 2023}{Hong Kong}

% USENIX stuff(?)
% Use the following at camera-ready time to suppress page numbers.
% Comment it out when you first submit the paper for review.
% \thispagestyle{empty}

% ACM Stuff
%%
%% The code below is generated by the tool at http://dl.acm.org/ccs.cfm.
%% Please copy and paste the code instead of the example below.
%%
\begin{CCSXML}
<ccs2012>
   <concept>
       <concept_id>10002978</concept_id>
       <concept_desc>Security and privacy</concept_desc>
       <concept_significance>500</concept_significance>
       </concept>
   <concept>
       <concept_id>10010520.10010553</concept_id>
       <concept_desc>Computer systems organization~Embedded and cyber-physical systems</concept_desc>
       <concept_significance>500</concept_significance>
       </concept>
 </ccs2012>
\end{CCSXML}

\ccsdesc[500]{Security and privacy}
\ccsdesc[500]{Computer systems organization~Embedded and cyber-physical systems}

%%
%% Keywords. The author(s) should pick words that accurately describe
%% the work being presented. Separate the keywords with commas.
\keywords{industrial control systems, programmable logic controllers, fuzzing}

%\received{29 March 2023}
%\received[revised]{9 July 2023}
%\received[accepted]{13 July 2023}

\begin{abstract}

Networked Programmable Logic Controllers (PLCs) are proprietary industrial devices utilized in critical infrastructure that execute control logic applications in complex proprietary runtime environments that provide standardized access to the hardware resources in the PLC. These control applications are programmed in domain-specific IEC 61131-3 languages, compiled into a proprietary binary format, and process data provided via industrial protocols. Control applications present an attack surface threatened by manipulated traffic. For example, remote code injection in a control application would directly allow to take over the PLC, threatening physical process damage and the safety of human operators. However, assessing the security of control applications is challenging due to domain-specific challenges and the limited availability of suitable methods. Network-based fuzzing is often the only way to test such devices but is inefficient without guidance from execution tracing.

This work presents the FieldFuzz framework that analyzes the security risks posed by %control applications for 
the Codesys runtime (used by over 400 devices from 80 industrial PLC vendors). FieldFuzz leverages efficient network-based fuzzing based on three main contributions: i) reverse-engineering enabled remote control of control applications and runtime components, 
ii) automated command discovery and status code extraction via network traffic and iii)  a monitoring setup to allow on-system tracing and coverage computation. We use FieldFuzz to run fuzzing campaigns, which uncover multiple vulnerabilities, leading to three reported CVE IDs. To study the cross-platform applicability of FieldFuzz, we reproduce the findings on a diverse set of Industrial Control System (ICS) devices, showing a significant improvement over the state-of-the-art.

\end{abstract}

\maketitle

\section{Introduction}\label{sec:introduction}

Industrial Control Systems (ICS) comprise critical infrastructure such as desalination plants, smart grids, transportation systems, and the nuclear sector. Digital control and communication in ICS are performed by proprietary Operational Technology (OT) devices which suffer from security challenges known to the IT domain (such as parsing bugs for traffic). In addition, OT devices commonly do not employ OS-level countermeasures that are standard practice in IT nowadays~\cite{abbasi2019challenges}. Exploiting such vulnerabilities could have catastrophic physical consequences, such as destroying equipment~\cite{proc:code_level_vulnerability}.%, and losing service for the population~\cite{proc:dark_and_gloomy}.

At the core of ICS, Programmable Logic Controllers (PLCs) process sensor readings while executing control logic to perform real-time control of physical processes. Engineers write control applications in programming languages defined by the IEC 61131-3 standard, and executed in proprietary runtime environments on the PLCs. The runtime environment, Integrated Development Environment (IDE), and compiler are proprietary software of the device manufacturer. The most significant Original Equipment Manufacturer (OEM) generic framework is the Codesys runtime, which is at the heart of $\approx$80 industrial device vendors ranging over 400 devices, including manufacturers such as Schneider Automation, Bosch Rexroth, WAGO, and Hitachi Europe~\cite{misc:codesys_device_list}, representing at least $\approx$20\% of the active PLCs worldwide~\cite{proc:icsfuzz}. 

Security threats to this ecosystem can be introduced, mainly via the control applications (through their handling of sensor data) and the diverse software components of runtimes. In addition, vendors can extend runtimes with their third-party libraries, exacerbating \emph{supply chain risks}, as the final runtime is a collection of components from various sources. The urgent/11 and ripple20 vulnerabilities~\cite{urgent11,ripple20} are recent examples of critical supply chain security issues in ICS devices. In both cases, low-level network traffic handling libraries contained bugs that allowed privileged remote code execution by the attacker, affecting millions of devices.

Given this, how could third parties (such as operators) test PLC devices and their control logic applications for (security) bugs? For analysis of proprietary IT software, fuzzing has been very successful in recent years. Unfortunately, due to domain-specific challenges, available fuzzers cannot directly be applied to PLC runtimes or their IEC 61131-3 control applications. In particular, PLC runtimes are complex stateful multithreaded applications that interact through proprietary protocols with the environment. Furthermore, control applications must be executed within the runtime, requiring new inputs (via memory-mapped I/O) in each scan cycle. To the best of our knowledge, ICSFuzz~\cite{proc:icsfuzz} is the only reported tool in the literature for fuzzing control applications. However, it suffers from significant drawbacks, such as losses in input delivery synchronization, slow fuzzing speed, manual crash monitoring, and only supporting a specific physical device (a WAGO PLC). Applying traditional fuzzers to this problem, on the other hand, like AFL++, also has limitations as such methods cannot control the runtime, often lack network capabilities, and cannot fuzz \emph{in situ}, i.e., correctly execute the bespoke control applications binaries used in ICS.

This work proposes a novel unified approach for vulnerability discovery throughout the computational stack of PLCs in the Codesys ecosystem (including their control applications). We realize this approach in the FieldFuzz framework, which is the first platform capable of fuzzing all components in a PLC, including the IEC 61131-3 control applications. It leverages reverse-engineered features of the proprietary Codesys platform and protocols to allow fuzzing of the runtime during its execution (in context), and the control applications in the runtime context, providing meaningful crashes. Furthermore, we designed Ghost, our system for obtaining accurate instruction-level coverage statistics, that is injected \emph{in situ} in the closed proprietary PLCs as a control application, allowing easy and multi-platform deployment. Our experiments discovered multiple vulnerabilities that were responsibly disclosed to corresponding device vendors. Three CVE IDs have been assigned to vulnerabilities discovered by FieldFuzz. %\footnote{We have asked the numbering authority and the vendor to anonymize the credits during the review phase.}

%To speed up our experiments, the target system can be run in a virtual machine, which allows us to perform snapshot-based fuzzing (e.g., fast resets of the targets). We also demonstrate fuzzing of targets that are running on native hardware.

% Our experiments resulted in the discovery of multiple vulnerabilities that were responsibly disclosed to corresponding device vendors. Currently, {\CVECOUNT} CVE IDs have been assigned to vulnerabilities discovered by {\FRAMEWORKNAME}\footnote{We have asked the numbering authority and the vendor to anonymize the credits during the review phase.
% %We expect more as we are analyzing more crashes uncovered by {\FRAMEWORKNAME}.
% }.

In summary, our main contributions are the following:
\begin{enumerate}[leftmargin=*, topsep=0pt,itemsep=-1ex,partopsep=1ex,parsep=1ex]

\item We reverse the Codesys runtime and its proprietary communication protocol to enable complete remote control over control application execution in the runtime, enabling fuzzing of the control applications and the runtime in their native execution environments, leading to three CVEs.

\item We demonstrate the feasibility and efficacy of our approach through our implementation and evaluation of FieldFuzz for both the control applications and the runtime. In addition, we introduce a local monitor called Ghost that provides further fuzzing feedback when injected into the target as a control application. As we leverage the standard communication protocols and runtime APIs, our approach allows cross-architecture fuzzing of diverse Codesys targets independent of the target platform and device vendor.

\item We demonstrate improvement over the state-of-the-art (i.e., ICSFuzz~\cite{proc:icsfuzz}) in control application fuzzing with higher performance, reliable scan cycle control, input delivery, monitoring, and breakpoint-based coverage feedback. FieldFuzz supports fuzzing runtime components that are inherently unreachable by ICSFuzz. In contrast to ICSFuzz, our approach applies to any target which supports Codesys (architecture independent).

\end{enumerate}

\section{Background and Prior Work}

\begin{figure}[tb]
\centering
\includegraphics[clip, trim=0cm 9.7cm 17.5cm 0cm, width=\columnwidth]{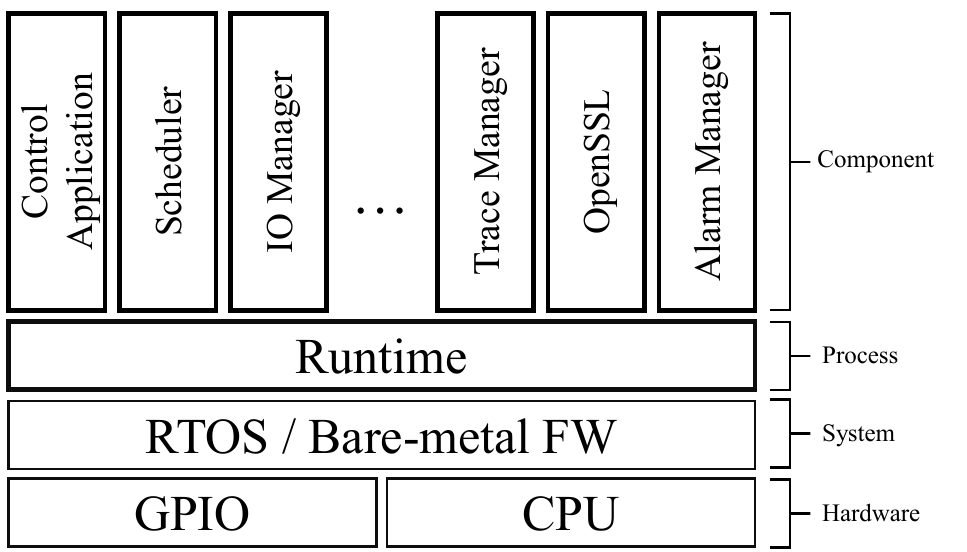}
\caption{PLC System stack. Blocks with thick borders belong to the proprietary codebase and require reverse engineering to facilitate fuzzing.}
\label{fig:runtime}
\end{figure}

\subsection{PLC Runtimes for Control Applications}
% While certain ICS platforms interpret the intermediate representation of the control program, others employ an IEC compiler to produce a control binary. These control binaries differ from known executable formats such as ELF or PE and cannot execute independently. Therefore, they are loaded and executed in the context of a PLC runtime, which controls every aspect of the binary execution (see Figure~\ref{fig:runtime}).
% % While they may share structural similarities such as header, program section, specific program entry point, linked libraries, these IEC binaries cannot execute independently.

% In general, PLC runtimes, such as the Codesys runtime that we focus on in this work are complex applications that encapsulate a wide array of functionality almost akin to a fully fledged operating system. As illustrated in Figure~\ref{fig:runtime}, these runtimes are typically deployed on top of minimal real-time operating systems, or bare-metal PLC firmware. In the first case, the runtime might rely on the functionality offered by the OS, while in the latter case, the runtime is normally incorporated as part of the firmware image. Regardless of the runtime deployment method, the runtime itself usually contains its own distinct but interconnected components that enable core functionality such as: executing the Control Application, I/O delivery, network communications, cryptographic functions, logging, event and exception handling, facilitating atomic operations etc.

Process engineers write control logic using IEC 61131-3 compliant programming languages (such as Ladder Logic) on an engineering workstation. While certain ICS platforms interpret the intermediate representation of the control program, others employ an IEC compiler to produce a compiled control binary. These control binaries differ from known executable formats such as ELF or PE, and cannot execute independently. Therefore, they are loaded and executed in complex PLC runtimes, which control every aspect of the binary execution (see Figure~\ref{fig:runtime}).

In general, PLC runtimes, such as the Codesys runtime that we focus on in this work, are complex applications that encapsulate a wide array of functionality almost akin to a fully-fledged operating system. As illustrated in Figure~\ref{fig:runtime}, these runtimes operate on top of minimal real-time operating systems or bare-metal PLC firmware. In the first case, the runtime might rely on the functionality offered by the OS, while in the latter case, the runtime is part of the firmware image. Regardless of the runtime deployment method, the runtime usually contains distinct but interconnected components that enable core functionality such as: executing the control application, I/O delivery, network communications, cryptographic functions, logging, event and exception handling, facilitating atomic operations, and more. Some components are solely purposed to interact with an open-source dependency (such as OpenSSL). In Figure~\ref{fig:runtime}, we mark the component responsible for communication with OpenSSL as part of the proprietary codebase. The reason behind this is that to provide such interaction, the runtime implements additional logic and wraps the core libraries into independent components of its proprietary format.
    
\subsection{Codesys Environment}

Codesys is a multi-platform software that includes the development system and the runtime for target ICS devices. The runtime is a collection of components with a modular architecture implemented as statically compiled and dynamically linked necessary libraries~\cite{codesys_brochure}, provided by Codesys, the device vendors, or open-source. At the same time, the increase in traditional vulnerabilities in control applications follows the evolution of the support of advanced external libraries \cite{misc:codesys-cve}. For example, consider \texttt{CVE-2020-6081}, which exploits code execution vulnerability in \texttt{PLC\_TASK} functionality of Codesys runtime 3.5.14.30, triggerable by a specially crafted network packet, enabling remote code execution. Furthermore, the runtime is also vulnerable to other classical vulnerabilities like out-of-bounds read (\texttt{CVE-2021-30194}), write (\texttt{CVE-2021-30193}), NULL pointer dereference (\texttt{CVE-2021-29241}), and more. 

\subsection{Prior Work}\label{sec:priorwork}
\noindent \textbf{Protocol fuzzing.} There is a considerable amount of work on fuzzing network protocols. For instance, AFLNET~\cite{proc:aflnet}, a greybox fuzzer fed with a mutated corpus of recorded network messages, utilizes state-feedback for guiding the fuzzer, and KiF~\cite{proc:kif} for fuzzing session initiation protocol. In addition, Pulsar is a stateful black-box fuzzing testing technique for proprietary network protocols that utilize automated reverse engineering and simulation~\cite{proc:pulsar}. There are ICS network protocol-specific solutions such as Peach*\cite{proc:peach_star}, a coverage-based improvement over the standard Peach fuzzer~\cite{misc:peach}. PropFuzz~\cite{proc:propfuzz} monitors the behavior of the controller along with the network connection to detect unexpectedly long jitters in the control process. Polar~\cite{art:polar} utilizes static and dynamic taint analysis to identify vulnerable operations with semantic aware mutation to improve the fuzzing procedure. Finally, ICS$^{3}$Fuzzer~\cite{proc:ics3fuzzer} is a framework for discovering implementation bugs in the supervisory software by fuzzing the network communication protocol employed to communicate with the field devices. It synchronizes the controls of the GUI operation and network communications to fuzz the entire supervisory software. It should be noted that ICS$^{3}$Fuzzer fuzzes the supervisor software and is not concerned with fuzzing PLC devices. While FieldFuzz can also do protocol fuzzing (as ICS communication is a component), it focuses on vulnerability discovery of \emph{any} component integrated into the PLC.

\noindent \textbf{Control application security.} Research on control application binaries focuses primarily on their safety verification. For instance, Canet et al.~\cite{proc:formal_verification_il} employ formal semantics and a model checking tool to verify rich behaviors and properties of Instruction List PLC programs. Other approaches detect malicious inputs to the PLC by verifying against temporal safety properties~\cite{proc:plc_mem_corruption}, and  monitoring violations in specifications~\cite{art:runtime_ics_specification}. In VetPLC, Zhang et al.~\cite{proc:plcvet} utilize static analysis for creating timed event graphs combined with invariant data traces to detect hidden safety violations. Guo et al.~\cite{proc:symplc} automatically translate control logic into C and perform symbolic execution to generate test cases. On the other hand, AttkFinder~\cite{proc:attkfinder} uses information flow-guided symbolic execution on an intermediate representation. Keliris et al.~\cite{art:icsref} reverse-engineered the Codesys v2.3 file format for control application binaries to propose an automated on-the-fly attack formulation. Similarly, CLIK~\cite{proc:clik} automatically modifies control logic executing on a PLC, and sends false data to the engineering software using captured network traffic.

The closest work on directly fuzzing industrial binaries is ICSFuzz~\cite{proc:icsfuzz}, a fuzzing framework for primarily fuzzing control applications. It supplies inputs to the control binary by overwriting the value at the memory-mapped GPIO and collects coverage metrics by overwriting NOP instructions. It detected multiple crashes for a synthetic control application binary dataset and a limited subset of runtime functions. However, as further discussed in Section \ref{sec:limitations}, it features numerous significant limitations.

\section{Fuzzing of Proprietary Industrial
Controllers}\label{sec:preliminaries}
%    ICS
%    
%    PLC
%    
%    famous attacks

%\subsection{Problem Formulation}
%    -Research Questions- 
%    \begin{itemize}
%        \item Q1: How to identify vulnerabilities in IEC 61131-3 applications?
%        \item Q2: How secure are the runtime (network) components?
%        \item Q3: How to assess without physical PLC at larger scale?
%    \end{itemize}
    
%    • Custom IEC 61131-3 applications are loaded into memory
%    • Extensive libraries and communication capabilities
%    • Programs are realtime and based on cyclic I/O supply
%    • How secure are these? How to identify vulnerabilities?
%    • How to scale the approach?

%\subsection{Threat Model}
%Examples of attack scenarios:
%    \begin{itemize}
%        \item PLC-to-PLC self-propagating I/O worm
%        \item  Rogue sensors to hijack the PLC  <-- new thing. <-- fuzz the IO drivers separately?
%        \item  PLC hijacks the HMI, spreads the attack
%        \item  (Automatic generation of unsafe states of IEC PRG)
%        \item  (Mass assessment of vulnerable IEC libs in the store)
%    \end{itemize}

\label{sec:problem}
% In this work,  we address the following overall problem: \emph{How to systematically and  efficiently identify (security) bugs in proprietary  PLC runtimes?} Achieving this goal will require us to address several research questions, which we will list next. Then, we summarize research and engineering challenges to achieving the overall goal. 
% Finally, we present a high-level overview of the proposed FieldFuzz solution.

This work addresses the following overall problem: \emph{How to systematically and efficiently identify (security) bugs in proprietary  PLC runtimes?} As listed below, achieving this goal requires addressing several research questions. Then, we summarize research and engineering challenges for achieving the overall goal. 

\subsection{Research Questions}
\begin{itemize}[leftmargin=*, topsep=0pt,itemsep=-1ex,partopsep=1ex,parsep=1ex]
%\item RQ1: What are key challenges to fuzz test for (security) bugs in proprietary industrial PLC devices? In particular, how to test the logic and data processing code in IEC applications in the PLC runtime?
% this assumes we don't know the challenges yet/ state them here. Would allow us to emphasize stronger why AFL/other approaches are insufficient.
\item RQ1: How to systematically (and efficiently) test control applications running in situ within a complex proprietary multi-threaded runtime such as the one used by Codesys? %In particular, how to ensure that the interactions of the control applications with the multi-threaded components within the runtime are appropriately captured?
%\item RQ1: How to systematically (and efficiently) test control applications running within a complex proprietary runtime such as the one used by Codesys? In particular, how to ensure that the interactions of the control applications with the multi-threaded components within the runtime are appropriately captured?
\item RQ2: How to guide the in situ testing of proprietary bare-metal runtimes without local access to the device?
\item RQ3: Is it possible to generalize the testing approach to be cross-platform?
% nils: RQ3 is a bit weak an a by-product out of our general approach.
\end{itemize}

\subsection{Research and Engineering Challenges}

Fuzzing the PLC stack is a challenging task for the following reasons: 
\begin{itemize}[leftmargin=*, topsep=0pt,itemsep=-1ex,partopsep=1ex,parsep=1ex] 

    \item PLC runtimes are typically closed-source proprietary software, necessitating black-box fuzzing and requiring extensive reverse engineering.
    \item Cross-platform PLC runtime rehosting is very difficult due to the complex interactions with hardware and peripherals. Even in the more straightforward case of IoT firmware, recent efforts have only achieved partial rehosting~\cite{proc:halucinator, proc:jetset}. At the same time, symbolic execution of the runtime and its control applications is also challenging, given the complexity of the binary, leaving fuzzing on the actual (proprietary) hardware as the only option.
    \item The PLC runtime runs as a root process (on some platforms, as a kernel module), and can be seen as a system-on-a-system since it takes over significant hardware resources (timers, I/Os, etc.) to ensure its real-time operation. Controlling it from a fuzzer is as challenging as controlling a full-blown operating system. 
\end{itemize}

\subsection{Threat Model and Assumptions}
% We assume the fuzzing of a PLC is conducted by the owner to identify security vulnerabilities present in the PLC runtime or the control applications executed within it. The PLC's runtime is available as a binary without debugging information. The owner can install additional control applications on the PLC (which we will leverage for the injection of the Ghost monitor).
 
% The goal is to identify vulnerabilities that could be exploited by an attacker who is able to monitor, intercept and modify network communication to the PLC, essentially performing a man-in-the-middle (MITM) attack (e.g., as in Stuxnet~\cite{art:stuxnet} and IRONGATE~\cite{art:irongate}). 
% Other similar research work such as TCP veto~\cite{proc:tcp_veto} and CLIK~\cite{proc:clik} employ the same assumption.

% The adversary can deliver malicious input to the target PLC through various approaches, such as:
% \begin{itemize}[leftmargin=*, topsep=0pt,itemsep=-1ex,partopsep=1ex,parsep=1ex]
%     \item Compromising a sensor that provides input to the PLC, explored in~\cite{proc:attacking_fieldbus}.
%     \item Intercepting and modifying network communication with the HMI terminal, which can allow data modification in the program~\cite{art:icsref} while supplying false data to the HMI.
% \end{itemize}

We assume that the owner of the PLC conducts its fuzzing to identify security vulnerabilities in the runtime or the control application executed within it. The PLC's runtime is available as a binary without debugging information, while the owner can install additional control applications on the PLC (leveraged for injecting the Ghost monitor). The goal is to identify vulnerabilities exploitable by an adversary who can monitor, intercept and modify sensor~\cite{proc:attacking_fieldbus} or network communication to the PLC~\cite{art:icsref}, essentially performing a man-in-the-middle (MITM) attack (e.g., as in Stuxnet~\cite{art:stuxnet} and IRONGATE~\cite{art:irongate}). 
Other similar research works employ the same assumption, such as TCP veto~\cite{proc:tcp_veto} and CLIK~\cite{proc:clik}.

%\begin{table}[]
%\centering
%\caption{CAPTION HERE}
%\label{table:related}
%\begin{tabular}{@{}rcccc@{}}
%\toprule
%& blackbox  & network & monitoring & scan cycle control   \\ \midrule
%Vanilla AFL & $\tilde{}$ & - & + & -  \\
%ICSFuzz  & + & - & - & -  \\
%Our work & + & + & + & +  \\ \bottomrule
%\end{tabular}
%\end{table}
\begin{table}[t]
\centering
\caption{Comparison of FieldFuzz with state-of-the-art for fuzzing control applications and runtime. \Circle, \LEFTcircle, and \CIRCLE represent non-existent, partial, and full support%, respectively
.}
\label{table:related}
\setlength\tabcolsep{3pt}
\resizebox{\linewidth}{!}{\begin{tabular}{|c|ccccc|ccccccc|cc|}
\hline
\multirow{2}{*}{Works} & \multicolumn{5}{c|}{Runtime} & \multicolumn{7}{c|}{Control Application} & \multicolumn{2}{c|}{Misc} \\ \cline{2-15} 
 & \multicolumn{1}{c|}{\rotatebox[origin=c]{90}{Support}} & \multicolumn{1}{c|}{\rotatebox[origin=c]{90}{In-context fuzzing}} & \multicolumn{1}{c|}{\rotatebox[origin=c]{90}{Input delivery}} & \multicolumn{1}{c|}{\rotatebox[origin=c]{90}{Coverage feedback}} & \rotatebox[origin=c]{90}{Crash monitoring} & \multicolumn{1}{c|}{\rotatebox[origin=c]{90}{Support}} & \multicolumn{1}{c|}{\rotatebox[origin=c]{90}{In-runtime fuzzing}} & \multicolumn{1}{c|}{\rotatebox[origin=c]{90}{Input delivery}} & \multicolumn{1}{c|}{\rotatebox[origin=c]{90}{Multi vendor support}} & \multicolumn{1}{c|}{\rotatebox[origin=c]{90}{Scan cycle control}} & \multicolumn{1}{c|}{\rotatebox[origin=c]{90}{Coverage feedback}} & \rotatebox[origin=c]{90}{Crash monitoring} & \multicolumn{1}{c|}{\rotatebox[origin=c]{90}{Input over network}} & \rotatebox[origin=c]{90}{Bare-metal fuzzing} \\ \hline
AFL++ \cite{AFLplusplus-Woot20} & \multicolumn{1}{c|}{\LEFTcircle} & \multicolumn{1}{c|}{\LEFTcircle} & \multicolumn{1}{c|}{\Circle} & \multicolumn{1}{c|}{\Circle} & \Circle & \multicolumn{1}{c|}{\Circle} & \multicolumn{1}{c|}{\Circle} & \multicolumn{1}{c|}{\Circle} & \multicolumn{1}{c|}{\Circle} & \multicolumn{1}{c|}{\Circle} & \multicolumn{1}{c|}{\Circle} & \Circle & \multicolumn{1}{c|}{\Circle}  & \Circle \\ \hline
ICSFuzz \cite{proc:icsfuzz} & \multicolumn{1}{c|}{\LEFTcircle} & \multicolumn{1}{c|}{\Circle} & \multicolumn{1}{c|}{\LEFTcircle} & \multicolumn{1}{c|}{\Circle} & \LEFTcircle & \multicolumn{1}{c|}{\CIRCLE} & \multicolumn{1}{c|}{\Circle} & \multicolumn{1}{c|}{\LEFTcircle} & \multicolumn{1}{c|}{\Circle} & \multicolumn{1}{c|}{\Circle} & \multicolumn{1}{c|}{\CIRCLE} & \LEFTcircle & \multicolumn{1}{c|}{\Circle}  & \Circle \\ \hline
FieldFuzz & \multicolumn{1}{c|}{\CIRCLE} & \multicolumn{1}{c|}{\CIRCLE} & \multicolumn{1}{c|}{\CIRCLE} & \multicolumn{1}{c|}{\LEFTcircle} & \CIRCLE & \multicolumn{1}{c|}{\CIRCLE} & \multicolumn{1}{c|}{\CIRCLE} & \multicolumn{1}{c|}{\CIRCLE} & \multicolumn{1}{c|}{\CIRCLE} & \multicolumn{1}{c|}{\CIRCLE} & \multicolumn{1}{c|}{\CIRCLE} & \CIRCLE & \multicolumn{1}{c|}{\CIRCLE}  & \CIRCLE \\ \hline
\end{tabular}}
\end{table}

% Removed: & \multicolumn{1}{c|}{\rotatebox[origin=c]{90}{Blackbox fuzzing}} 

\subsection{Limitations of existing methods}\label{sec:limitations}

While other approaches can potentially be applied to fuzzing control logic applications, these either fail at addressing the challenges mentioned above or feature significant limitations that hinder efforts towards efficient and reliable in situ fuzzing (see Table~\ref{table:related}). Two such approaches are AFL++~\cite{AFLplusplus-Woot20}, a generic general-purpose fuzzer, and ICSFuzz~\cite{proc:icsfuzz}, explicitly built for fuzzing control applications.

Vanilla AFL++ cannot fuzz control applications for several reasons. Firstly, control binaries are compiled into bespoke formats meant for execution within the context of their runtimes; thus, AFL++ cannot execute them outside their runtime environment. Moreover, AFL++ cannot control the runtime or the scan cycle, emulate input delivery from peripherals, monitor the control application and runtime for crashes, and fails to perform in-runtime fuzzing for such programs. Additionally, AFL++ generally requires shell access, making it incompatible with industrial controllers where the runtime is distributed as a bare-metal firmware image with no underlying OS or kernel module. Finally, AFL++ lacks network fuzzing capabilities, which prevents fuzzing proprietary PLCs.

ICSFuzz, on the other hand, can fuzz control applications running on PLCs but features significant limitations that hamper fuzzing efficiency. First, it utilizes the KBUS subsystem for input delivery to the control application, necessitating a physical PLC, and making it non-scalable. Due to a lack of scan cycle synchronization, it periodically drops fuzzing inputs and is slow. Furthermore, ICSFuzz lacks automation for observing the state of the control program and involves manual crash monitoring. Finally, it also performs limited stateless fuzzing of the shared library functions of the Codesys runtime on WAGO PLC, discovering some crashes. However, due to the stateless and out-of-context nature of the fuzzing, it misses vulnerabilities requiring the execution context of the runtime.

AFL++ and ICSFuzz cannot directly interact with the runtime by calling its specific functions, nor can they maintain the runtime state. Consequently this also prevents them from fuzzing the entirety of the runtime, limiting their reach and capabilities. FieldFuzz, on the other hand, is created explicitly for fuzzing in the ICS environment and addresses all the above limitations. Its methodology is designed to be particularly generic, and control application fuzzing is just an instance of component fuzzing, as was the case with protocol fuzzing discussed in Section \ref{sec:priorwork}.

\section{FieldFuzz}\label{sec:FRD}
%As discussed in the problem formulation (Section \ref{sec:problem}), 

%IEC application binaries are not plain executables that can be fuzzed independently. Instead, they have to be interacted while being loaded and executed by the runtime. Therefore, in order to fuzz the IEC binary, we need to gain the control over the runtime to deliver desired inputs and receive crash feedback. 

%In this section we present an overview of the proposed approach to address our research questions. Consequently we discuss details on FieldFuzz and Ghost, which enable stateful system-level fuzzing and coverage tracking. 

\begin{figure}[tb]
\centering
\includegraphics[clip, trim=0.5cm 12cm 10cm 0cm, width=0.85\columnwidth]{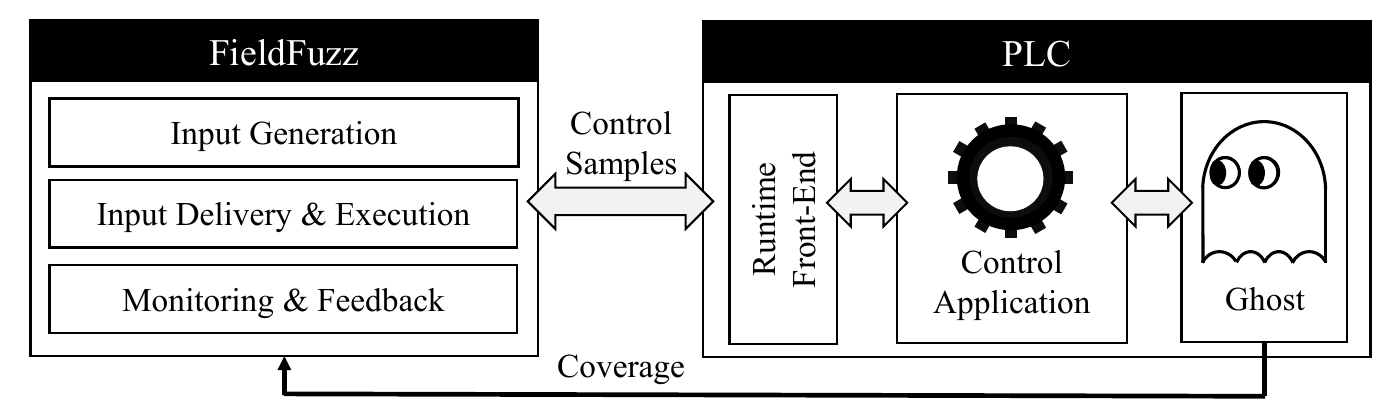}
\caption{Overview of FieldFuzz. It controls and tests the target control application via the reversed runtime protocol while Ghost obtains additional coverage data.}
\label{fig:overview}
\vspace{-0.2cm}
\end{figure}

% {\color{red} WIP: Break this into a general overview, explain how these answer RQs. Keep points from the text below that was originally in Section III}
%We propose to address the research questions and reach our goal through a novel custom fuzzing framework, which allows \emph{in situ} fuzzing of the IEC application (and other components) \emph{within} the PLC runtime. To allow this fuzzing to run on black-box proprietary PLC hardware, we fuzz via the (to be reverse-engineered) network interface of the runtime (addressing RQ1). To allow more efficient fuzzing, we propose to enrich the fuzzing feedback with a custom application that is injected into the target via the IEC application mechanism, we call this custom application Ghost (addressing RQ2). Utilizing the standardized (proprietary) network interface should also allow cross-platform fuzzing, which we will experimentally validate (addressing RQ3). Figure~\ref{fig:overview} provides a high level overview of the proposed system.

\begin{figure*}[t]
\centering
\includegraphics[clip, trim=0.2cm 5.7cm 0cm 5.2cm, width=1.0\linewidth, keepaspectratio]{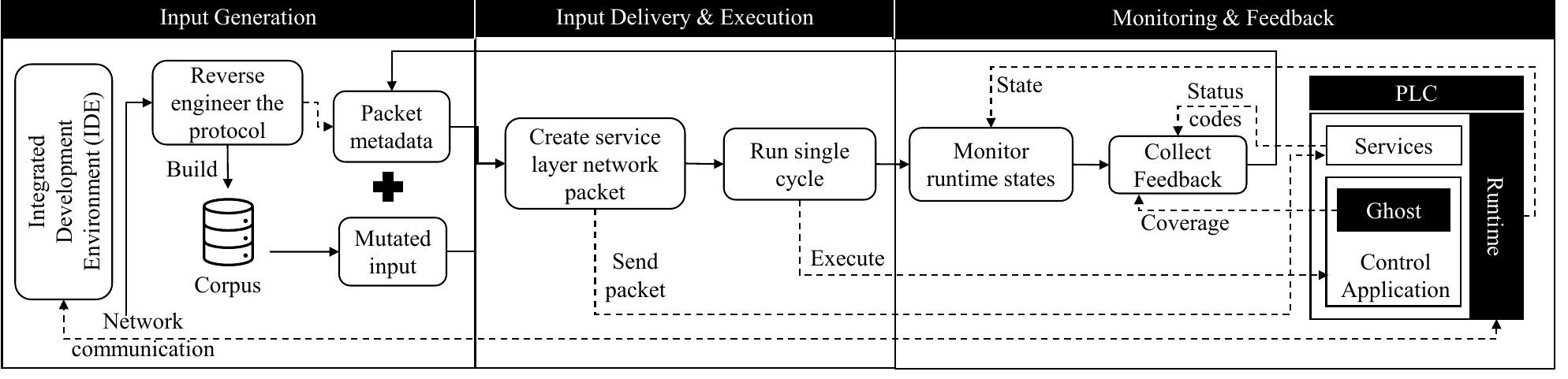}
\caption{FieldFuzz methodology overview for fuzzing control applications and the runtime.}
\label{fig:methodology}
\vspace{-0.15cm}
\end{figure*}

We address the research questions and reach our goals through a novel custom fuzzing framework, which allows \emph{in situ} fuzzing of the control application (and other components) \emph{within} the context of the PLC runtime. To allow this fuzzing to run on black-box proprietary PLC hardware, we fuzz via the network interface of the runtime (addressing RQ1). To allow for more efficient fuzzing, we inject a custom application, dubbed Ghost, into the target via the control application loading mechanism (addressing RQ2), providing FieldFuzz with fuzzing feedback. Utilizing the standardized (proprietary) network interface should also allow cross-platform fuzzing, which we validate experimentally (addressing RQ3). Figure~\ref{fig:overview} provides a high-level overview of the proposed system.

\subsection{Overview}

As illustrated in more detail in Figure \ref{fig:methodology}, our proposed methodology consists of 3 interconnected stages. 1) In the Input Generation stage, we leverage the IDE to communicate with the runtime, reverse engineer its Codesys v3 communication protocol to understand packet metadata, and build an input corpus for fuzzing. This input corpus is further mutated based on the feedback from the Monitoring \& Feedback stage. 2) Since FieldFuzz utilizes the network interface to provide inputs to the control application and the runtime, the Input Delivery \& Execution stage encapsulates the mutated input into the proper packet structure. This packet is parsable by the runtime and is sent over the network. It fuzzes the control application by delivering inputs and controlling its execution by sending service requests to the runtime for single scan cycle execution. 3) In the Monitoring \& Feedback stage, FieldFuzz communicates with the runtime services to monitor the runtime state and collect feedback. Additionally, it receives coverage feedback from our Ghost monitor, injected into the runtime via the control application loading mechanism. Finally, the complete feedback is relayed to the Input Generation stage to generate future inputs and guide the fuzzing efforts.

\begin{table*}[t]
\caption{Commands associated with components of our interest at the service layer. In bold, are the commands replicated by FieldFuzz that are crucial for the fuzzing routines. Additional markings correspond to the fuzzing stage: ` I ': Initialization, ` D ': Input delivery, ` E ': Execution control, ` M ': Monitoring.}
\label{table:commands}
\resizebox{\linewidth}{!}{\begin{tabular}{|c|c|c|c|c|c|c|c|c|c|c|c|}
\hline
\textbf{Cmp}                                                                                                                   & \textbf{Command}                                  & \textbf{ID} & \textbf{Cmp}                                                                         & \textbf{Command}                              & \textbf{ID} & \textbf{Cmp}                                                                         & \textbf{Command}                               & \textbf{ID} & \textbf{Cmp}                                                                                                                          & \textbf{Command}                           & \textbf{ID} \\ \hline
\multirow{10}{*}{\rotatebox[origin=c]{90}{CmpDevice {[}0x01{]}}}                                        & \textbf{GetTargetIdent$^{I, M}$} & 0x01        & \multirow{13}{*}{\rotatebox[origin=c]{90}{CmpApp {[}0x02{]}}} & DeleteApp                                     & 0x04        & \multirow{13}{*}{\rotatebox[origin=c]{90}{CmpApp {[}0x02{]}}} & \textbf{ReadAppList$^{I, M}$} & 0x18        & \multirow{10}{*}{\rotatebox[origin=c]{90}{CmpApp {[}0x02{]}}}                                                  & ReadProjectInfo                            & 0x31        \\ \cline{2-3} \cline{5-6} \cline{8-9} \cline{11-12} 
                                                                                                                               & \textbf{Login$^{I}$}             & 0x02        &                                                                                      & Download                                      & 0x05        &                                                                                      & SetNextStatement                               & 0x19        &                                                                                                                                       & DefineFlow                                 & 0x32        \\ \cline{2-3} \cline{5-6} \cline{8-9} \cline{11-12} 
                                                                                                                               & \textbf{Logout$^{I}$}            & 0x03        &                                                                                      & OnlineChange                                  & 0x06        &                                                                                      & ReleaseForceList                               & 0x20        &                                                                                                                                       & ReadFlowValues                             & 0x33        \\ \cline{2-3} \cline{5-6} \cline{8-9} \cline{11-12} 
                                                                                                                               & \textbf{SessionCreate$^{I}$}     & 0x0A        &                                                                                      & DeviceDownload                                & 0x07        &                                                                                      & UploadForceList                                & 0x21        &                                                                                                                                       & DownloadEncrypted                          & 0x34        \\ \cline{2-3} \cline{5-6} \cline{8-9} \cline{11-12} 
                                                                                                                               & ResetOrigin                                       & 0x04        &                                                                                      & CreateDevApp                                  & 0x08        &                                                                                      & \textbf{SingleCycle$^{D}$}    & 0x22        &                                                                                                                                       & ReadAppContent                             & 0x35        \\ \cline{2-3} \cline{5-6} \cline{8-9} \cline{11-12} 
                                                                                                                               & EchoService                                       & 0x05        &                                                                                      & \textbf{Start$^{I}$}         & 0x10        &                                                                                      & CreateBootProject                              & 0x23        &                                                                                                                                       & SaveRetains                                & 0x36        \\ \cline{2-3} \cline{5-6} \cline{8-9} \cline{11-12} 
                                                                                                                               & SetOperatingMode                                  & 0x06        &                                                                                      & Stop                                          & 0x11        &                                                                                      & ReInitApp                                      & 0x24        &                                                                                                                                       & RestoreRetains                             & 0x37        \\ \cline{2-3} \cline{5-6} \cline{8-9} \cline{11-12} 
                                                                                                                               & GetOperatingMode                                  & 0x07        &                                                                                      & \textbf{Reset$^{I, E}$}      & 0x12        &                                                                                      & ReadAppStateList                               & 0x25        &                                                                                                                                       & \textbf{GetAreaAddress$^{D}$}                             & 0x38        \\ \cline{2-3} \cline{5-6} \cline{8-9} \cline{11-12} 
                                                                                                                               & InteractiveLogin                                  & 0x08        &                                                                                      & SetBP         & 0x13        &                                                                                      & LoadBootApp                                    & 0x26        &                                                                                                                                       & LeaveExecpointsActive                      & 0x39        \\ \cline{2-3} \cline{5-6} \cline{8-9} \cline{11-12} 
                                                                                                                               & RenameNode                                        & 0x09        &                                                                                      & \textbf{ReadStatus$^{M}$} & 0x14        &                                                                                      & RegisterBootApp                                & 0x27        &                                                                                                                                       & ClaimExecpointsForApp                      & 0x40        \\ \cline{1-3} \cline{5-6} \cline{8-12} 
\multirow{3}{*}{\rotatebox[origin=c]{90}{\begin{tabular}[c]{@{}c@{}}CmpApp \quad \\  {[}0x02{]}\end{tabular}}} & \textbf{Login$^{I}$}             & 0x01        &                                                                                      & DeleteBP      & 0x15        &                                                                                      & CheckFileConsistency                           & 0x28        & \multirow{2}{*}{\begin{tabular}[c]{@{}c@{}}CmpMonitor2\\ {[}0x1B{]}\end{tabular}} & Read                                       & 0x01        \\ \cline{2-3} \cline{5-6} \cline{8-9} \cline{11-12} 
                                                                                                                               & \textbf{Logout$^{I}$}            & 0x02        &                                                                                      & ReadCallStack                                 & 0x16        &                                                                                      & \textbf{ReadAppInfo$^{I}$}    & 0x29        &                                                                                                                                       & \textbf{Write$^{D}$}      & 0x02        \\ \cline{2-3} \cline{5-6} \cline{8-12} 
                                                                                                                               & CreateApp                                         & 0x03        &                                                                                      & GetAreaOffset & 0x17        &                                                                                      & DownloadCompact                                & 0x30        & \begin{tabular}[c]{@{}c@{}}CmpPlcShell\\ {[}0x11{]}\end{tabular}               & \textbf{Execute$^{M, E}$} & 0x01        \\ \hline
 \end{tabular}}
\end{table*}
%  I for initialization, D for input delivery,  E for execution control, M for monitoring, C for code coverage
% \begin{tablenotes}
%     \centering
%     \small
%     \item ` I ': Initialization, ` D ': Input delivery, ` E ': Execution control, ` M ': Monitoring, ` C ': Code coverage
% \end{tablenotes}

\subsection{Input Generation}
FieldFuzz leverages the proprietary communication protocol utilized by Codesys to communicate with the runtime for enabling fuzzing over the network. This protocol facilitates hierarchical device-to-device communication between the engineering software (IDE) and the target devices (PLC, HMI touch panels, Gateways). The flexible nature of the protocol allows its routing in a single industrial network with diverse segments of Ethernet, CAN, Serial, Sercos, and other media. Without loss of generality, FieldFuzz connects to field devices with TCP over Ethernet.

To gain sufficient knowledge for implementing FieldFuzz, we developed a complete Wireshark dissector for the Codesys v3 communication protocol, written in Lua. 
Using the dissector's parsing and filtering capabilities on the captured traffic, combined with manual reverse engineering of the runtime, and published information~\cite{misc:kaspersky_codesys}, we extracted all the information needed to develop FieldFuzz. 
 This also allowed us to collect a corpus of valid interactions with components, and communication patterns required for stateful (maintaining session and runtime state) fuzzing.
% Clarification: a rough estimate of the time required for reverse engineering
In Section~{\ref{sec:Methodology}}, we discuss and estimate the manual steps that are required to fuzz the runtime components.

The runtime utilizes a proprietary network stack to facilitate network communication between nodes\footnote{Detailed information about the network stack is available in~\cite{misc:kaspersky_codesys}.},
% . The Codesys v3 network stack consists of four layers:
with four layers:

\noindent \textbf{Block layer.} This layer is responsible for communication over the physical interfaces. It processes a verification number and the total number of bytes in the packet. It then transfers the rest of the packet to the Datagram layer.

\noindent \textbf{Datagram layer.} This layer detects Codesys nodes in the network and routes the packets appropriately. It parses the verification number, hop information structure, a \textit{service\_group\_id} identifying the destination service, a packet length field, the sender and receiver addresses, and optional padding.

\noindent \textbf{Channel layer.} This layer ensures synchronized communication, integrity verification, and delivery acknowledgment. Communications rely on opening, maintaining, and closing communication channels. Beyond data, packets contain an ID that indicates the desired functionality concerning channel management, a flag field, a \textit{channel\_id} for identifying the open channel, and packet metadata (e.g., checksums, acknowledgment IDs, etc.).

\noindent \textbf{Service layer.} This layer is responsible for querying the requested service and transmitting the operating settings. Among others, the message in this layer contains the \textit{service\_group\_id} field, which refers to the unique ID that the runtime uses internally to identify available services. The message contains \textit{command\_id} to indicate the target command within a specific service. Finally, the message also features session, content, and protocol metadata fields.

We identify the invariant and variable fields in the packet structure based on the packet captures collected from the network communication between the Codesys IDE and the runtime. \emph{Invariant} fields are the metadata information required by FieldFuzz to construct the input delivery packet, such as service group (identifying the requested service), command ID (identifying the requested runtime command), and more. Such fields do not change values between similar categories of requests. On the other hand, \emph{variable} fields whose value changes in the communication are the potential fuzzing inputs. For instance, if a user forces a value change for a variable in the control application, the varying fields will be limited to the user-specified forced input value, address offsets for the variable in the control application memory space, the size of the variable, session identifiers, and more. However, the invariant fields, such as service and command IDs, will remain consistent, making them easily identifiable. Using such a procedure, we build metadata information for requesting particular runtime services, and identify fuzzing input fields. Typically, Service Group IDs (2 bytes) for the vendor-added components specific to a runtime variant are enumerable and within a dedicated range beginning from \texttt{0x100}, along with the Command IDs (2 bytes). FieldFuzz sequentially enumerates the 2 bytes of the Command ID and reads the returned status code to determine existing commands. Some of the commands validate the device-level and application-level session or both. The standard status codes indicate which commands do not exist for the enumerated Service Group ID. We save the valid tuples as component interfaces for fuzzing.

Table \ref{table:commands} presents a subset of commands associated with a select few runtime components and is crucial for our fuzzing routines. The commands identified and replicated by FieldFuzz are denoted in bold and marked according to their utilization in different stages of our fuzzing methodology. While these are the commands whose input format is pre-programmed and known to FieldFuzz, we show later in Section \ref{sec:results} that FieldFuzz allows interaction with any component of choice that is reachable from the network by specifying an ID tuple for routing, and the corpus to produce the input.

\subsection{Input Delivery \& Execution}
In literature, bare-metal IO modules are often utilized as the primary communication method for PLCs~\cite{proc:icsfuzz, art:malware_physics}. However, such an approach is not scalable and is often specific to the model or series of the PLC. Another approach is to utilize the Modbus protocol from the Fieldbus family of network communication protocols. However, such an approach would require the project to include the Modbus client object to receive read and write commands, requiring explicit declarations of exports in the control project, usually disabled by default. Commonly, the HMI displays and modifies inputs to a program using the OPC Unified Architecture (OPC UA) protocol. However, OPC UA support is not present in all Codesys distributions and requires a license; otherwise, the OPC UA server shuts down after 2 hours. Furthermore, using symbolic access for input delivery requires implicit mapping of the global variable list to the project, allowing the operator workstation to access chosen variables by name.

The abovementioned approaches require explicit configuration modification in the control project to support specific Fieldbus protocols and symbolic variables. To address this problem, FieldFuzz utilizes a universal approach and does not require project modification while enabling read and write to any variable regardless of its type and visibility scope. It utilizes \emph{tags}, a nested binary structure to send requests, command payloads, and replies to the service layer of the runtime as defined in its proprietary network stack. Each tag starts with its ID: \textit{tag\_id}, which often corresponds to the type of payload or status code, the \textit{tag\_size}, \textit{tag\_data}, and some \textit{additional\_data}. The tag also contains \textit{data\_size}; the size of the supplied data, \textit{write\_value}; the value to be written, and \textit{write\_offset}, which is the offset of the target variable from the start of the data section.

FieldFuzz initiates communication with the runtime over the network using its proprietary protocol while the runtime listens on multiple ports for connection requests (TCP 1217, 11740, and UDP 1740 to 1743). Depending on the requested service layer command, it establishes sessions in at least two stages to reliably invoke commands in the target components: the device, and the application login stage. FieldFuzz utilizes the \texttt{CmpDevice} Login command to retrieve a device-level session handle from the runtime, which the runtime associates with the channel in its mapping tables. It then retrieves a list of loaded applications from \texttt{CmpApp} component. If the application is loaded on the PLC, FieldFuzz can open an additional session via the \texttt{CmpApp} Login command to obtain the application session ID and its handle. The session information is maintained to perform stateful operations with the runtime. Finally, it implements a keepalive mechanism to keep the channel active despite the timeout imposed by the runtime.

\noindent \textbf{Input delivery to the runtime components.} To deliver inputs to the control application and other runtime components, FieldFuzz first creates a (\textit{service\_group\_id}, \textit{command\_id}) tuple to identify the appropriate component for routing the input. Next, using the recursive tag encoder algorithm~\cite{misc:kaspersky_codesys_2}, it builds a binary tag structure to place the value into a corresponding field based on the corpus of pre-known Tag IDs and their value formats. The resulting binary structure is included in the body of a Service Layer packet. Next, FieldFuzz calculates a CRC32 checksum for this data, constructs a packet header, and encapsulates it with all the remaining layers. 

For delivering the input, FieldFuzz sends this prepared packet containing the fuzzing input to the runtime over the active channel. The Block Layer of the runtime receives the packet, processes it, and passes it to a component that parses the \textit{service\_group\_id} from the packet header and routes the packet to the target component. Finally, the communication handler of the target component performs sanity checks for the data format and passes the input binary stream to the function that implements the corresponding \textit{command\_id}.

It should be noted that since FieldFuzz provides fuzzing inputs to the control application and the runtime components over the network, it needs to maintain the session information relating to the state of the corresponding runtime component. It maintains the state of \textit{blk\_id} and \textit{ack\_id} counters required for generating valid subsequent packet headers. It also stores the application session ID, and its handle. Furthermore, identifying known states is essential for uncovering vulnerabilities since the components of the runtime are interconnected such that they call functions and access structures belonging to each other. This process provides inputs to target components, enabling stateful component fuzzing.

\noindent \textbf{Input delivery to the control application.} Delivering inputs to the control application involves several additional routines. FieldFuzz achieves this by performing a write operation over the network to the control application memory segment. Typically, the runtime separates data and code (as compiled instructions) into separate memory segments, referred to as Area0 and Area3, respectively. FieldFuzz reads and writes variables using relative offset addresses in the Area0 memory segment. For this, it constructs short bytecode programs as inputs for the \texttt{CmpMonitor2} component, using a set of 58 opcodes specific to the runtime. Depending on the syntax, the opcodes can read inputs from the interpreter stack and receive up to 4 inline arguments. For implementation, we extract the opcode table from the reverse-engineered runtime binary and map their names and format using a relevant decompiled dynamic library loaded by the Codesys IDE on the operator workstation. The interpreter performs various checks and returns the status code in the first byte of the tag (e.g., wrong pointer, buffer overrun, and more). FieldFuzz detects a successful read operation with the presence of tag \texttt{0x40} in the reply, and a failed attempt by the presence of tag \texttt{0x41}. In contrast, the write operation requires a more complex bytecode program, and returns no data in reply for a successful attempt. Nevertheless, FieldFuzz can still verify the write (input delivery) operation by detecting the presence of tag \texttt{0x41}.

\subsection{Monitoring \& Feedback}
Understanding code coverage is essential to assessing the effectiveness of our fuzzing campaigns. However, this is particularly challenging since the Codesys runtime is proprietary software. As a result, we are forced to utilize blackbox fuzzing on the binary level without access to the source code, which is not straightforward since the runtime employs a binary packing mechanism as an anti-tampering measure, hindering our efforts to locate relevant code blocks corresponding to the components. Additional analysis of the dumped memory regions of the runtime shows that the runtime binary is stripped of its debug symbols during compilation, further increasing the difficulty of locating interesting components for fuzzing.

To address the challenges of visibility into the fuzzed execution of the runtime components, FieldFuzz utilizes a combination of status codes, and a Ghost monitor loaded using the standard control program loading mechanism, to enable and infer code coverage.

\noindent \textbf{Status code for execution feedback.} When a component executes a requested command, it returns a value to the callee, indicating the completion status of the requested operation. These internal identifiers are called status codes and differ from Service Group IDs mapped to the components. 

As a code path coverage mechanism, FieldFuzz watches for the status codes returned by the different layers of the Codesys v3 network protocol stack and the output of various runtime components, maintaining the status code sequence during the fuzzing session. Based on the status code sequence changes resulting from mutated inputs, it understands and differentiates the execution path inside the component function. The status codes differ as the fuzzing input traverses the lower network layers. Essentially, a different part of the target function returns a different status code due to a change in the execution path. Therefore, whenever fuzzing uncovers a new execution path (by observing the status code), FieldFuzz adds it to the list of known states and initiates a new fuzzing campaign. For example, a reply with only Channel Layer status codes indicates a failure to reach the Service Layer for processing. The status codes can determine whether the command reached the target function or failed due to the lack of authentication, wrong Command ID, or incorrect payload format. 
% For a detailed review of common status codes used by runtime components, refer to Table~\ref{table:error_codes} in the Appendix.
Some of the notable status codes include those corresponding to missing tags (\texttt{L7TagMissing}), and non-existing command groups (\texttt{L7UnknownCmdGrp}) or commands (\texttt{L7UnknownCmd}). 

To retrieve the status of the execution of the control application binary, FieldFuzz surveys the \texttt{CmpApp} and \texttt{CmpPlcShell} components. Upon exception, a core dump and crash log from \texttt{CmpLog} are retrieved from the controller remotely for further investigation. Furthermore, to detect crashes of the runtime, it constantly monitors the latency in the channel and the consistency of the Block Layer counter.

\begin{figure}[t]
\centering
\includegraphics[clip, trim=0cm 9.92cm 19.13cm 0cm, width=\columnwidth]{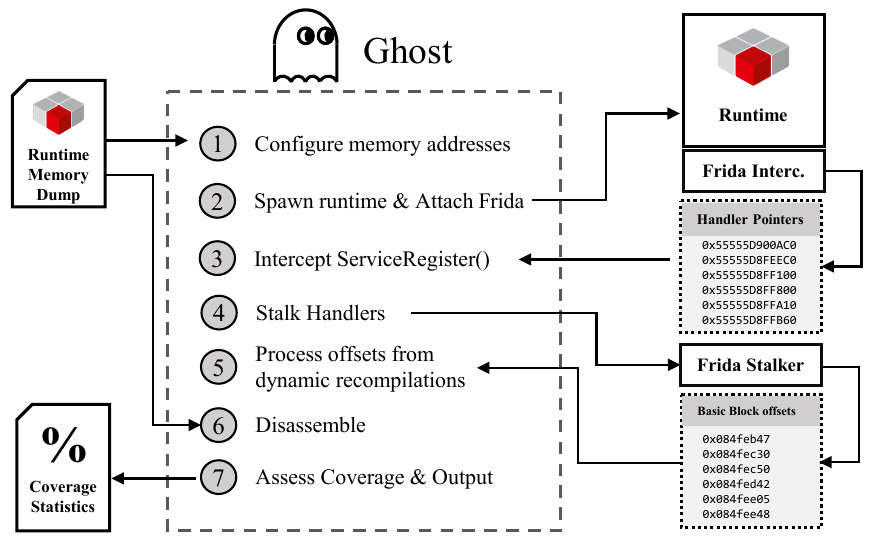}
\caption{The steps involved in the operation of Ghost.}
\label{fig:ghost}
\vspace{-0.2cm}
\end{figure}

\label{sec:ghost}
\noindent \textbf{Ghost for code coverage feedback.} During the runtime initialization phase, component setup functions make calls to a function used to register runtime components. These calls include memory pointers to the entry points of the runtime components. In order to capture these memory pointers, we leverage Frida\cite{misc:frida_website}, a dynamic instrumentation toolkit widely used for reverse engineering. Using Frida, we dynamically intercept calls to the registration function and locate the memory pointers to the component entry point. Then, using these pointers as starting points, we analyze the components and locate all the memory segments relevant to their commands. The use of Frida in our toolchain does not hamper our approach on more limited systems, as it can be compiled for a variety of architectures and utilized through its C API.

Accurate assembly instruction-level coverage requires instrumentation and monitoring of the spawned runtime threads for handling command invocation. Unfortunately, anti-tampering mechanisms hamper attempts at static instrumentation, so we utilize Frida to monitor for calls to component functions. When a call is intercepted, we launch Frida's code tracing engine that dynamically recompiles assembly code and traces the executing thread. By getting feedback from the recompilation process, we can keep track of executed basic blocks using their offsets in memory.

To automate this process, we introduce Ghost, our tool for dynamic black box instrumentation of the Codesys runtime to obtain accurate instruction-level coverage statistics (see Figure~\ref{fig:ghost}). Ghost is initially configured with access to the runtime memory dump and the relevant memory segments of the target components. Upon execution, it spawns the runtime and immediately independently attaches a set of scripts to capture all runtime component setup calls to avoid Codesys anti-tampering measures that execute on startup. Having captured the pointers to component functions, Ghost utilizes Frida to capture calls to them and consequently traces the runtime threads executing component code. While fuzzing with FieldFuzz, Ghost receives the notifications of dynamic recompilation events with their memory offsets, indicating the current execution of a basic block. Ghost then disassembles the relevant memory segment and keeps track of the executed assembly instructions. Upon exit, it saves the context information for component functions, a log of the dynamic recompilation events, and extensive instruction-level coverage information.

\subsection{Protocol Dissector}
To gain sufficient knowledge for implementing FieldFuzz and to support the fuzzing campaigns, we have developed a complete Wireshark dissector for the Codesys v3 communication protocol.
The full-featured dissector for Codesys v3 protocol enables the use of Wireshark with advanced filter expressions and custom columns for every aspect of the Codesys v3 stack. 
Based on the knowledge gained in this work, we also provide the dissector with symbols to translate the packet data into human-readable component, command and field names. 
The dissector significantly streamlines the process of input corpus collection, crash investigation and exploit development. It is implemented using the Wireshark Lua API and provides the following capabilities:

\begin{enumerate}
    \item Live decoding of the network traffic capture (IDE-to-PLC, PLC-to-PLC, Gateways, HMIs)
    \item Analysis of the pre-recorded dump files
    \item Detection of the Codesys network stack on non-standard ports using magic constants
    \item Using advanced filter expressions for all layers of the Codesys v3 stack
    \item Using custom columns for all layers of the Codesys v3 stack
    \item Choosing between the use of human-readable symbols and raw values for filters and column data or combining both
    \item Direct access to the Layer 7 binary tag tree and recursive extraction of the payloads
\end{enumerate}

\section{Fuzzing Campaigns}\label{sec:Methodology}
 %\subsection{Fuzzing Industrial Runtime}
% To demonstrate that we addressed RQ1 and RQ2, we now apply our framework to fuzz the Codesys runtime and control application binaries.

To address RQ1 and RQ2, we apply the FieldFuzz framework to fuzz the Codesys runtime and control application binaries.
An overview of the binaries is provided in Sections~\ref{sec:crossarch} and~\ref{sec:results}, respectively.

\subsection{Fuzzing Setup}
\label{subsec:fuzzing_setup}

\begin{figure}[t]
\centering
\includegraphics[clip, trim=1.5cm 8.1cm 15.3cm 5.3cm, width=\columnwidth]{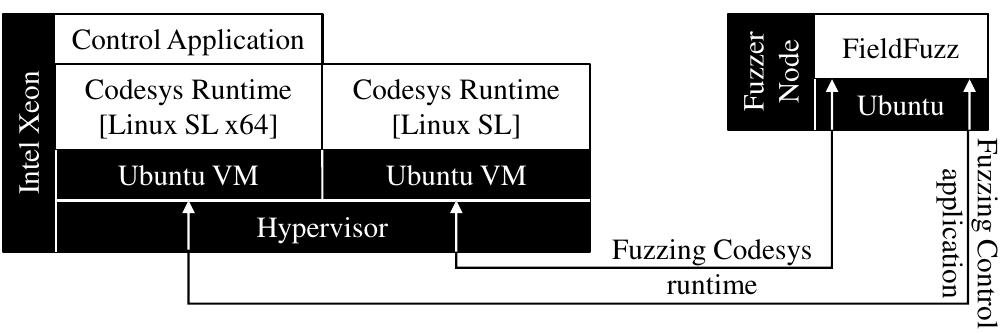}
\caption{Experimental setup for fuzzing control applications and the runtime.}
\label{fig:experiment_setup}
\vspace{-0.1cm}
\end{figure}

Figure~\ref{fig:experiment_setup} shows the basic experimental setup for fuzzing the Codesys runtime and control application binaries. We utilize two virtual machines with the Codesys runtime variant for Linux, which run on the Intel Xeon-based hypervisor server. The runtime includes a standard init.d wrapper that facilitates the automatic runtime restart after a crash caused by FieldFuzz. In addition, we disable the system-wide address space layout randomization (ASLR) on these virtualized nodes to simplify the crash investigation. 

\noindent \textbf{Scalability options.} It should be noted that FieldFuzz supports directly fuzzing physical devices, however, these are optional. FieldFuzz can utilize multiple virtual machines to scale the fuzzing setup. On the other hand, another potential possibility to scale the experiment in a single VM is to utilize the ability of the channel layer of the runtime to handle multiple channels simultaneously, a functionality supported by FieldFuzz. While FieldFuzz primarily uses TCP over Ethernet, it can fuzz any node in the plant by relaying the packets through other devices, including the nodes of the network that are not reachable by Ethernet, such as those connected by serial interface or CAN bus.

% % Clarification: a rough estimate of the time required for reverse engineering and clarify how much manual work was required to find the previously-unknown vulnerabilities.
\noindent \textbf{Estimation of manual efforts.} Assuming that the operator is already familiar with the FieldFuzz Framework and the Wireshark dissector, the manual efforts required to derive an input corpus and setup fuzzing for a typical standard component of the runtime can be roughly estimated as one hour of work.
%In the case of a non-standard component for which the network communication can not be captured, additional time for manual reverse engineering may be required.
%However, due to the structured nature of the components and the consistent pattern in their network handler function implementation, this should only take several additional hours of work.
If no traffic capture is available, an input corpus can be bootstrapped through limited reverse-engineering of the component packet handler in the runtime. In our experience during reverse engineering the runtime, we found that packet handler code is quite structured and deriving sufficient information from the binary to produce correct messages can be done in a few hours.

\begin{table}[t]
\centering
\caption{Components and service groups at the service layer.}
\label{table:components}
\resizebox{0.7\columnwidth}{!}{\begin{tabular}{|c|c|c|c|}
\hline
\textbf{Component}    & \textbf{ID} & \textbf{Component}  & \textbf{ID} \\ \hline
CmpAlarmManager & 0x18        & CmpLog        & 0x05        \\ \hline
CmpApp          & 0x02        & CmpMonitor    & 0x03        \\ \hline
CmpAppBP        & 0x12        & CmpMonitor2   & 0x1B        \\ \hline
CmpAppForce     & 0x13        & CmpOpenSSL    & 0x22        \\ \hline
CmpCodeMeter    & 0x1D        & CmpSettings   & 0x06        \\ \hline
CmpCoreDump     & 0x1F        & CmpTraceMgr   & 0x0F        \\ \hline
CmpDevice       & 0x10        & CmpUserMgr    & 0x0C        \\ \hline
CmpFileTransfer & 0x08        & CmpVisuServer & 0x04        \\ \hline
CmpIecVarAccess & 0x09        & CmpPlcShell      & 0x11        \\ \hline
CmpIoMgr        & 0x0B        & SysEthernet   & 0x07        \\ \hline
\end{tabular}}
\vspace{-0.2cm}
\end{table}

\noindent \textbf{Identifying fuzzing runtime targets.} As discussed earlier, the runtime is a collection of components (including the component responsible for executing control binaries), so fuzzing the runtime implies interaction with the components responsible for its functionality. However, despite the runtime having a single generic codebase, its actual builds can significantly vary based on the target architecture, vendor modifications, and hardware platform constraints. Therefore, the first step is to create a complete list of all instantiated components. To achieve that, we start by extracting the interfaces of components reachable from the network. The \emph{component interfaces} are defined as the tuple: (Service Group, Command). First, we identify the runtime components loaded by the particular runtime from the boot log of the device and its device description file used by the IDE. Next, we identify the Service Group IDs of the loaded components. For the generic set of components developed by Codesys, we get this information from the decompiled libraries of the IDE and the captured network communication. Table \ref{table:components} presents a subset of components in our target runtime variant.

\noindent \textbf{Fuzzing inputs.} To collect a dataset of valid inputs, we trigger commands with the Codesys IDE and capture its communication with the runtime. We extract Service layer payloads and decode the nested binary tag structure using our Wireshark dissector. We then save the tag IDs, structure, and valid payloads for each (Service Group, Command) tuple. Next, we determine the packet fields derived from the session identifier to identify the variant fields for maintaining state information. Finally, we save these inputs as seeds with the identified format for the current runtime distribution. After creating an initial corpus of input seeds, we utilize python bindings for libradamsa~\cite{misc:radamsa} mutators ported from AFL++~\cite{AFLplusplus-Woot20} to mutate the byte payload. 
% Also, there is no discussion about fuzzed data variation (invalid data, unexpected data, random data). They can mention their data generation techniques briefly.
We generate byte blocks of variable length to pass them inside the tags without mutating the tag ID and preserving the remaining structure of the corpus. To prevent fragmentation of the packets that carry the generated payloads, we calculate the maximum length constraint for all tags in order to not exceed the fragmentation threshold value of the runtime (512 bytes in the default installation) cumulatively, including the layer headers. We apply the set of techniques provided by the mutation engine of libradamsa in default execution mode (including Bit and Byte Flipping, Block Swapping, Value Substitution, Byte expansion) to form unexpected data inside the byte blocks. It should be emphasized that mutations are semi-random and are not currently guided by coverage (which is a topic of future work).
FieldFuzz then delivers the input to the runtime over the network.

% === Kostas notes: ===
% - Why code coverage
% - Challenges: closed source => binary, packed binary => memory unpacking & dumping, striped symbols => identification via Interceptor, short lived threads + basic blocks => GumEvents + Stalker
% - Solution: Ghost
% - Deployment: Use Codesys APIs in ST to deploy

\noindent \textbf{Code coverage.} The Ghost monitor calculates coverage on the component, handler, and command level (see Section~\ref{sec:ghost}). On the component level, coverage includes all the memory segments associated with a component, complete with the command code and its entry-point function. Entry function coverage indicates the percentage of instructions executed within the entry function for running the various component commands. Finally, command coverage is limited to the memory segments relevant to each command.

\noindent \textbf{Deployment.} Performing the above steps and preparing to run Ghost to obtain coverage information can be particularly challenging on PLCs that do not offer shell access. We work around this by exploiting functionality readily available within Codesys by embedding the relevant binaries into the control application project such that it is downloaded onto the local file system while loading the control application binary on the PLC. It also loads the Ghost deployment script written in Structured Text employing the native \texttt{SysFileCopy} API (available in the \texttt{SysFile} library) to relocate the binary files in the file system. Since the runtime runs with root privileges, we utilize \texttt{SysFileOpen} and \texttt{SysFileWrite} to modify the \texttt{CODESYSControl.cfg} runtime configuration file by appending \texttt{[SysProcess]Command=AllowAll} to it for enabling arbitrary shell command execution. Having done this, we restart the runtime through the IDE or reboot the target PLC to force the configuration changes. Upon reboot, we use \texttt{SysProcessExecuteCommand2} (part of the \texttt{SysProcess} library) to perform the necessary setup and run Ghost with root privileges.

% \begin{table}
% \centering
% \caption{Code coverage recorded while fuzzing commands belonging to the Codesys Trace Manager component. %Total component coverage is 44.72\%. Service handler coverage is 54.76\%.
% }
% \label{tbl:coverage}
% \begin{tabular}{c!{\vrule width \lightrulewidth}c!{\vrule width \lightrulewidth}c} 
% \toprule
% \textbf{Command} & \textbf{Command ID} & \textbf{Command Coverage}  \\ 
% \midrule
% RecordAdd        & 0x0D                & 95.14\%                    \\
% PacketCreate     & 0x02                & 56.17\%                    \\
% PacketClose      & 0x06                & 97.56\%                    \\
% PacketComplete   & 0x04                & 97.56\%                    \\
% PacketStart      & 0x0A                & 97.56\%                    \\
% PacketRead       & 0X07                & 26.98\%                    \\
% PacketStop       & 0X0B                & 97.56\%                    \\
% PacketGetConfig  & 0X0F                & 71.54\%                    \\
% PacketOpen       & 0x05                & 92.5\%                     \\
% PacketReadList   & 0X01                & 90.48\%                    \\
% \bottomrule
% \end{tabular}
% \end{table}

\begin{table}
\centering
\caption{Code coverage recorded while fuzzing commands belonging to the Codesys Trace Manager component.}
\label{tbl:coverage}
\resizebox{0.75\columnwidth}{!}{
\begin{tabular}{|c|c|c|} 
\hline
\textbf{Command} & \textbf{Command ID} & \textbf{Command Coverage}  \\ 
\hline
RecordAdd        & 0x0D                & 95.14\%                    \\ 
\hline
PacketCreate     & 0x02                & 56.17\%                    \\ 
\hline
PacketClose      & 0x06                & 97.56\%                    \\ 
\hline
PacketComplete   & 0x04                & 97.56\%                    \\ 
\hline
PacketStart      & 0x0A                & 97.56\%                    \\ 
\hline
PacketRead       & 0X07                & 26.98\%                    \\ 
\hline
PacketStop       & 0X0B                & 97.56\%                    \\ 
\hline
PacketGetConfig  & 0X0F                & 71.54\%                    \\ 
\hline
PacketOpen       & 0x05                & 92.5\%                     \\ 
\hline
PacketReadList   & 0X01                & 90.48\%                    \\
\hline
\end{tabular}
}
\vspace{-0.2cm}
\end{table}

As a proof of concept, we run fuzzing campaigns on functions of three standard components: \texttt{CmpTraceMgr}, \texttt{CmpPlcShell}, and \texttt{CmpDevice}. 
To derive the input corpus and initialize the campaigns, approximately one hour of manual work was required per each fuzzed component.
FieldFuzz was able to uncover a variety of crashes, which we then analyzed, focusing on uncovering vulnerabilities. For brevity, we provide extensive discussions for one CVE and shorter discussions for the others. It should be noted that the CVEs found through our FieldFuzz campaigns could not have been discovered using other state of the art fuzzers like ICSFuzz. This is because the fuzzing process requires interacting with runtime components which cannot be reached by these fuzzers.

\subsection{Fuzzing CmpTraceMgr Component (CVE-2022-22514)}

%Below is the analysis of CVE-2021-34604
% The \texttt{CmpTraceMgr} component consists of eight critical operations triggered by the service layer commands in sequence. This default component is available in most full-featured distributions of the runtime. It is a backend for the Trace program organization unit object that is used in control application projects for recording and visualizing variable trends in the physical process. %Although none of the IEC projects running on the tested devices had this object, the component was still reachable and responded to the commands. 
% Here, the \texttt{recordAdd} operation causes \texttt{SEGFAULT} for two reasons:
% \begin{itemize}[leftmargin=*, topsep=0pt,itemsep=-1ex,partopsep=1ex,parsep=1ex]
%     \item As the command is sent out-of-order, the component enters an unexpected state where it operates on a pointer to a structure of a packet object that is never correctly initialized.
%     \item The offset calculation into this non-existent structure is based on the value supplied by {\FRAMEWORKNAME}. The attacker can influence it by controlling this offset, leading the runtime to perform \texttt{mov} operations on invalid memory addresses.
% \end{itemize}

\noindent \textbf{Setup.} To investigate the crash, we use FieldFuzz to generate a standalone exploit from a template. It connects with a SoftPLC (VM) node with full-featured debugging capabilities and sends the service layer payload to the Codesys runtime. To observe critical runtime errors, we enable core dumps and disable the error handling behavior of the \texttt{SysExcept} component that intercepts POSIX signals from the runtime for internal interpretation. We modify \texttt{CODESYSControl.cfg} to disable the internal exception handler and instruct the runtime to append the logs to a file with a permissive log filtering mask. To record core dumps, we launch the runtime binary (\texttt{codesyscontrol.bin}) as a standalone process outside its \texttt{init.d} service wrapper and provide it with a \texttt{-d} flag for detailed logging.

\noindent \textbf{Crash analysis.} FieldFuzz reported a crash for service group \texttt{0x0F}, command ID \texttt{0x0D}, belonging to the \texttt{CmpTraceMgr} component. Our examination of the original pcap file with the dissector revealed that the \texttt{recordAdd} (\texttt{0x0D}) command (with a 148 bytes payload) causes the crash. This payload incorporates three levels of nested binary tags. 

The \texttt{CmpTraceMgr} component consists of eight critical operations which are triggered by the service layer commands in sequence. This default component is available in most full-featured distributions of the runtime. It is a backend for the Trace program organization unit object used in control application projects for recording and visualizing variable trends in the physical process. Here, the \texttt{recordAdd} operation causes \texttt{SEGFAULT} because when the command is sent out-of-order, the component enters an unexpected state operating on a pointer to a structure of a packet object that is never correctly initialized. The offset calculation into this non-existent structure is controlled by FieldFuzz input. An adversary can exploit this offset, forcing the runtime to perform \texttt{mov} operations on invalid memory addresses.

\noindent \textbf{Call stack investigation.} At least 12 network stack functions handle the packet before it finally reaches the function related to \texttt{CmpSrv}, which is the top component of the network stack. Finally, \texttt{CmpSrv} calls an exported hook function of \texttt{CmpTraceMgr}, which acts as a handler for all service layer commands for the service group \texttt{0x0F}. The hook function extracts the command ID from the packet header and jumps into the condition based on command \texttt{0x0D}. Functions imported from the \texttt{CmpBinTagUtil} component parse the fuzzing input, and decode the 17 binary tags, including tag \texttt{0x40} which influences calculations of a memory address offset, the value for which is controlled by FieldFuzz. Consequently, a \texttt{SIGSEGV} occurs in the command handler function for the \texttt{recordAdd} command, caused by a \texttt{mov} instruction attempting to access the nonexistent memory address. The corrupted memory offset is from a structure that stores a tracing packet derived from the value supplied by FieldFuzz.

% \noindent \textbf{Status codes.}
% The component changes its returned status codes based on the multiple execution path conditions. The \texttt{recordAdd} function does several sanity checks for the supplied value. A reply containing the tag \texttt{0xFF7F} with a status code \texttt{0x02} was caused by the payloads in tag \texttt{0x40} that are outside the expected range, such as \texttt{0x0} and \texttt{0xFFFFFFFF}. This prevents the crash as the read operation is not reached.
% Another state of the component, indicated by the returned status code \texttt{0x11}, neither causes a crash nor forms a successful trace packet processing result. In this case, the payload falls into the allowed range and passes the entry checks of the \texttt{recordAdd} function. The read operation is reached, resulting in a handled exception due to returned empty packet data.

% \noindent \textbf{Coverage.} Table \ref{tbl:coverage} presents the coverage reported by Ghost during fuzzing campaigns on the \texttt{CmpTraceMgr} component. Our fuzzing strategy obtained high coverage for the majority of the component commands. In the cases of \texttt{PacketCreate} and \texttt{PacketRead}, upon examining the relevant instruction blocks, low coverage can be attributed to a lack of generation of stateful sequence command events during input mutation. Employing stateful input mutation strategies can improve the coverage for such command \cite{proc:banks2006snooze, proc:program_adaptive,proc:pavfuzz}. However, input mutation strategies are orthogonal to our work and will be explored in future research.

\noindent \textbf{Status code example.} The component changes its returned status codes based on the multiple execution path conditions. The \texttt{recordAdd} function does several sanity checks for the supplied value. For example, a reply containing the tag \texttt{0xFF7F} with a status code \texttt{0x02} is caused by the payloads in tag \texttt{0x40} that are outside the expected range, such as \texttt{0x0} and \texttt{0xFFFFFFFF}, preventing the crash by sanitizing inputs before reaching the vulnerable instruction. Another state of the component, indicated by the returned status code \texttt{0x11}, neither causes a crash nor forms a successful trace packet processing result. In this case, the payload falls into the allowed range and passes the elementary entry checks of the \texttt{recordAdd} function. This input reaches the vulnerable read operation; however, the exception is handled, and the component returns a packet without any data.

\noindent \textbf{Coverage.} Table \ref{tbl:coverage} presents the coverage reported by Ghost during fuzzing campaigns on the \texttt{CmpTraceMgr} component. Our fuzzing strategy obtained high coverage for the majority of the component commands. In the cases of \texttt{PacketCreate} and \texttt{PacketRead}, upon examining the relevant instruction blocks, low coverage can be attributed to a lack of generation of stateful sequence command events during input mutation. Employing stateful input mutation strategies can improve the coverage for such command \cite{proc:banks2006snooze, proc:program_adaptive,proc:pavfuzz}. However, input mutation strategies are orthogonal to our work and will be explored in future research.

This vulnerability has been reported to the vendor, and was assigned CVE-2022-22514.

%\noindent \textbf{Summary.} There are two fundamental causes for this vulnerability. First, the command reaches the component in an unexpected state such that it does not correctly initialize the memory segment of the trace packet. Second, the runtime derives the memory offset for the read operation from the tag \texttt{0x40} which the adversary controls.
    
%\begin{itemize}

\subsection{Fuzzing CmpDevice Component (CVE-2022-22508)}

\texttt{CmpDevice} is an essential component responsible for the authentication and network discovery of the PLC. It uses the \texttt{SetNodeName} (\texttt{0x09}) command for changing an identification string employed for in-network discovery and initiating a connection with the PLC. Unfortunately, a specially crafted packet sent to the runtime prevents the IDE from communicating with the PLC, resulting in a connection error. Moreover, this issue is persistent even across reboots because the payload from the network packet ends up in a persistent runtime configuration file and keeps restoring upon device boot. The vulnerability was reported to the vendor, and \texttt{CVE-2022-22508} was assigned.

The runtime becomes unresponsive due to a specially crafted packet sent to Service Group \texttt{0x01} (\texttt{CmpDevice}), command \texttt{0x09} (\texttt{SetNodeName}), with tag \texttt{0x58}, and a long bytestring consisting of non-printable characters as a service layer payload. This crafted bytestring is not sanitized properly by \texttt{CmpDevice} before being passed to the local \texttt{SysTarget} component (potentially vendor-specific, we tested on official Codesys distributions), and then stored permanently in the \texttt{NodeNameUnicode} property field. The device is not accessible even after a reboot because the node identifier is appended to the \texttt{CODESYSControl.cfg} configuration file as a new record. \texttt{CmpSettings} processes this file which is then consumed by \texttt{SysTarget}. The connecting client attempts to perform device discovery through the channel layer and calls \texttt{CmpDevice} again to perform \texttt{GetTargetIdent} and \texttt{CreateSession} commands. System log messages suggest that \texttt{CmpNameServiceServer}, a channel layer component that exports its functions to \texttt{CmpRouter} and implements a Codesys-specific naming system protocol~\cite{misc:kaspersky_codesys}, processes the bytestring. Consequently, the device fails to respond to further scan requests, and the dynamic libraries of the Codesys IDE raise several exceptions. Manual removal of the \texttt{SysTarget} section from the runtime configuration file restores the operational state of the device after a reboot.

\subsection{Fuzzing CmpPlcShell Component (CVE-2022-22507)}
\label{app:a22}

\texttt{CmpPlcShell} is a default built-in component that fetches information from the device, such as firmware revision and system load. It can also perform system diagnostics of the device by sending string commands of a particular format. An adversary can trigger a segmentation fault, crashing all the runtime threads by sending a specially crafted payload from the Codesys v3 network stack. The vulnerability was reported to the vendor, and \texttt{CVE-2022-22507} was assigned.

The main command body is passed inside tag \texttt{0x10}, while an additional tag \texttt{0x12} is required by some commands for handling the arguments. FieldFuzz detects the crash for the tag \texttt{0x12} because the runtime performs a memory read operation outside valid memory boundaries. By sending a sequence of packets, it is possible to force the runtime to perform memory access operations and enumerate the valid address range with the offset increments. As the offset grows in each operation by an internal loop, an unhandled \texttt{SIGSEGV} fault occurs once the operation exceeds valid memory boundaries.

\section{Cross-Architecture Generalization}
\label{sec:crossarch}

\begin{table}[t]
\centering
\caption{Codesys runtime binaries for different targets.}
\label{table:codesys_variations}
\resizebox{0.7\columnwidth}{!}{\begin{tabular}{|c|c|c|c|}
\hline
\textbf{Device}  & \textbf{Arch} & \textbf{Size (MB)} & \textbf{Packed} \\ \hline
WAGO PFC200      & arm32                 & 4.6                & \xmark          \\ \hline
BeagleBone Black & arm32                 & 5.8                & \cmark          \\ \hline
Linux SoftPLC    & x64                   & 9.7                & \cmark          \\ \hline
Raspberry Pi     & arm32                 & 5.5                & \cmark          \\ \hline
SIMATIC IOT2000  & x32                   & 6.4                & \xmark          \\ \hline
emPC-A/iMX6      & arm32                 & 6                  & \cmark          \\ \hline
Windows RTE      & x32                   & 103.9              & \cmark          \\ \hline
\end{tabular}}
\vspace{-0.2cm}
\end{table}

To address RQ3, we discuss the cross-architecture generalization of our approach. While the runtime has a single generic codebase, specifics for each target platform and architecture are reflected in different build variants. For instance, on platforms driven by the VxWorks real-time operating system (RTOS), the entire Codesys runtime is shipped as a kernel module. The embedded bare-metal runtime variant has a much smaller set of components but implements more complex system components to interact with the hardware. In more modern ICS devices powered by RTLinux (such as WAGO Touch Panel 600 series or WAGO PFC200 PLC), the runtime operates as root in the userspace and reuses resources provided by the OS, such as network sockets, timers, and file descriptors. 

% Clarification: the binaries they used to unpack (evaluation of binaries is crucial to the performance evaluation of the fuzzer). 
\noindent \textbf{Runtime binaries.}
As shown in Table~\ref{table:codesys_variations}, the size of the runtime binaries varies across various architectures from 4.6 MB to 103.9 MB. This is because the number of components and shared libraries linked statically or dynamically differs across different variants. In addition, some binaries are packed, involving license management and anti-tampering mechanisms. Our primary distribution of choice in this work (Codesys Control for Linux x64) employs a packing mechanism that we reverse dynamically by dumping the memory segments of the live process. On the other hand, the runtime variant for Windows devices (Codesys Control RTE x32) includes custom renamed and encrypted sections. From the section names and the protection function, we have noticed that these are managed by Wibu-Systems CodeMeter protection software~\cite{misc:codemeter_website}, which has also been used by Siemens and Rockwell~\cite{misc:codemeter_uscert}.

\begin{figure*}[t]
\centering
\includegraphics[clip, trim=0.5cm 14cm 2.5cm 1.2cm, width=\linewidth]{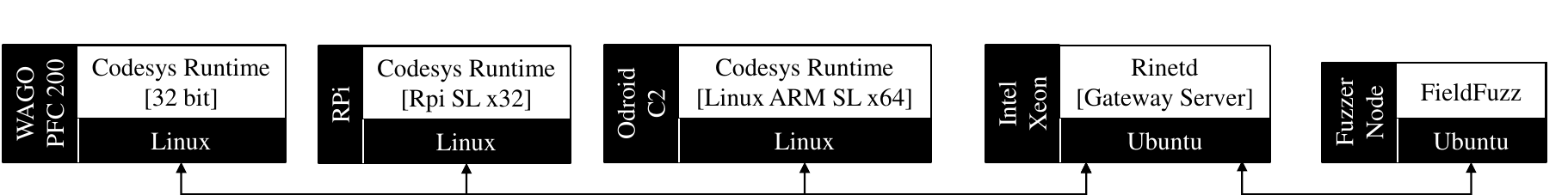}
\caption{Experimental setup for cross-architecture validation of the fuzzing results.}
\label{fig:gen_setup}
\vspace{-0.2cm}
\end{figure*}

\texttt{SysMem}, \texttt{SysSocket}, and \texttt{SysCom} are some critical hardware-dependent components in the runtime. At some point in the execution path, other components rely on the exported functions provided by these lower-level system components, which can influence the behavior of the crashes. Therefore, to assess the applicability of our findings, we test the attacks against different runtime variants by employing physical devices, as shown in Figure~\ref{fig:gen_setup}. We utilize a replayer node that initiates communication with the Gateway. The latter forwards the communications to multiple platforms in parallel. In this case, the Intel Xeon server acts as a VM hypervisor and the Gateway to WAGO PFC200, Raspberry Pi 4, and Odroid C2. This setup enables FieldFuzz to quickly test the same payload across multiple architectures and variants of the runtime. We observe the differences in the behavior of the crashes to adjust the input payload and port it between architectures. As a proof of concept, we replay the fuzzing inputs for crashing the \texttt{CmpTraceMgr} component (\texttt{CVE-2022-22514)}), which is available on all of the tested devices. The payload corresponding to the input is passed through the tag \texttt{0x40} and is four bytes long. On an x86 system with ASLR disabled, the crash input causes a \texttt{SIGSEGV}. However, with ASLR enabled, replaying the same value does not lead to a stable \texttt{SIGSEGV} because the resulting offset in the \texttt{recordAdd} function in most of the trials points to an unexpected but valid memory address. As a result, the command function of the component returns a status code (\texttt{0x11}), preventing the crash. On an x64 system, even enumerating the entire 4-byte range did not cause a crash. Nevertheless, such runtime variants accept longer payloads (8 bytes), eventually leading to the crash. The payload behaves identically on Intel and ARM, causing crashes on both the VM and physical devices; it only differs between 32 and 64 bit architectures of the target device.

\section{Fuzzing Control Application Binaries}\label{sec:results}
 %\textbf{Cross-platform comparison}
% Compare with icsfuzz or more
%\begin{itemize}
%\item more input methods, not only WAGO kbus
%\item Presence of scan cycle **D: beat the scan cycle? 1us
%\item  Performance
%\item Multi-platform
%\item Automated Monitoring, investigation
%\end{itemize}

\begin{table*}[t]
%\caption{Performance comparison with ICSFuzz. Speed for \FRAMEWORKNAME~is when using 1 VM; Speed for ICSFuzz is when using 1~PLC.}
\caption{Performance comparison of FieldFuzz against ICSFuzz. The execution speed metrics for FieldFuzz and ICSFuzz correspond to employing one VM and physical PLC, respectively. Markings: 1 - BeagleBone (ARM), 2 - Linux x64 (Intel), 3 - Wago PFC100 (ARM).}
%\vspace{-0.1in}
\label{table:icsfuzz_comparison}
\centering
\resizebox{0.8\linewidth}{!}{

\begin{tabular}{|c|ccc|ccc|ccc|ccc|}
\hline
\multirow{3}{*}{\textbf{\begin{tabular}[c]{@{}c@{}}Control\\ Applications\end{tabular}}} & \multicolumn{3}{c|}{\textbf{\begin{tabular}[c]{@{}c@{}}Execution Speed\\ (inputs/sec)\end{tabular}}}             & \multicolumn{3}{c|}{\textbf{\begin{tabular}[c]{@{}c@{}}First Crash\\ (seconds)\end{tabular}}}                    & \multicolumn{3}{c|}{\textbf{\begin{tabular}[c]{@{}c@{}}First Crash\\ (inputs)\end{tabular}}}                     & \multicolumn{3}{c|}{\textbf{Crashes}}                                                                            \\ \cline{2-13} 
                                                                                         & \multicolumn{2}{c|}{\textbf{{\FRAMEWORKNAME}}}                                    & \textbf{\cite{proc:icsfuzz}} & \multicolumn{2}{c|}{\textbf{{\FRAMEWORKNAME}}}                                    & \textbf{\cite{proc:icsfuzz}} & \multicolumn{2}{c|}{\textbf{{\FRAMEWORKNAME}}}                                    & \textbf{\cite{proc:icsfuzz}} & \multicolumn{2}{c|}{\textbf{{\FRAMEWORKNAME}}}                                    & \textbf{\cite{proc:icsfuzz}} \\ \cline{2-13} 
                                                                                         & \multicolumn{1}{c|}{\textbf{A8$^{1}$}} & \multicolumn{1}{c|}{\textbf{x64$^{2}$}} & \textbf{A8$^{3}$}           & \multicolumn{1}{c|}{\textbf{A8$^{1}$}} & \multicolumn{1}{c|}{\textbf{x64$^{2}$}} & \textbf{A8$^{3}$}           & \multicolumn{1}{c|}{\textbf{A8$^{1}$}} & \multicolumn{1}{c|}{\textbf{x64$^{2}$}} & \textbf{A8$^{3}$}           & \multicolumn{1}{c|}{\textbf{A8$^{1}$}} & \multicolumn{1}{c|}{\textbf{x64$^{2}$}} & \textbf{A8$^{3}$}           \\ \hline
bf\_mcpy\_1                                                                              & \multicolumn{1}{c|}{294.6}              & \multicolumn{1}{c|}{593}                & 70.88                        & \multicolumn{1}{c|}{0.014}              & \multicolumn{1}{c|}{0.25}               & 234                          & \multicolumn{1}{c|}{6}                  & \multicolumn{1}{c|}{148}                & 15270                        & \multicolumn{1}{c|}{8289}               & \multicolumn{1}{c|}{7876}               & 32                           \\ \hline
bf\_mcpy\_6                                                                              & \multicolumn{1}{c|}{286.4}              & \multicolumn{1}{c|}{642.4}              & 64.2                         & \multicolumn{1}{c|}{0.071}              & \multicolumn{1}{c|}{1.43}               & 188                          & \multicolumn{1}{c|}{22}                 & \multicolumn{1}{c|}{898}                & 12172                        & \multicolumn{1}{c|}{290}                & \multicolumn{1}{c|}{2384}               & 21                           \\ \hline
bf\_mcpy\_8                                                                              & \multicolumn{1}{c|}{244.6}              & \multicolumn{1}{c|}{645.6}              & 66.06                        & \multicolumn{1}{c|}{21.34}              & \multicolumn{1}{c|}{7.08}               & 279                          & \multicolumn{1}{c|}{5124}               & \multicolumn{1}{c|}{4566}               & 18216                        & \multicolumn{1}{c|}{145}                & \multicolumn{1}{c|}{359}                & 17                           \\ \hline
bf\_mcpy\_12                                                                             & \multicolumn{1}{c|}{320.6}              & \multicolumn{1}{c|}{526.2}              & 62.11                        & \multicolumn{1}{c|}{0.584}              & \multicolumn{1}{c|}{1.95}               & 426                          & \multicolumn{1}{c|}{181}                & \multicolumn{1}{c|}{999}                & 26645                        & \multicolumn{1}{c|}{847}                & \multicolumn{1}{c|}{977}                & 9                            \\ \hline
bf\_mset\_1                                                                              & \multicolumn{1}{c|}{223.3}              & \multicolumn{1}{c|}{560.6}              & 64.56                        & \multicolumn{1}{c|}{0.027}              & \multicolumn{1}{c|}{0.04}               & 208                          & \multicolumn{1}{c|}{8}                  & \multicolumn{1}{c|}{22}                 & 13441                        & \multicolumn{1}{c|}{22200}              & \multicolumn{1}{c|}{18105}              & 21                           \\ \hline
bf\_mset\_3                                                                              & \multicolumn{1}{c|}{268.6}              & \multicolumn{1}{c|}{571.2}              & 62.68                        & \multicolumn{1}{c|}{0.063}              & \multicolumn{1}{c|}{0.03}               & 174                          & \multicolumn{1}{c|}{8}                  & \multicolumn{1}{c|}{17}                 & 10906                        & \multicolumn{1}{c|}{21723}              & \multicolumn{1}{c|}{16085}              & 24                           \\ \hline
bf\_mset\_5                                                                              & \multicolumn{1}{c|}{289.3}              & \multicolumn{1}{c|}{503.2}              & 68.8                         & \multicolumn{1}{c|}{0.008}              & \multicolumn{1}{c|}{0.56}               & 254                          & \multicolumn{1}{c|}{2}                  & \multicolumn{1}{c|}{281}                & 17554                        & \multicolumn{1}{c|}{4447}               & \multicolumn{1}{c|}{4373}               & 16                           \\ \hline
bf\_mset\_9                                                                              & \multicolumn{1}{c|}{314.3}              & \multicolumn{1}{c|}{584.8}              & 69.76                        & \multicolumn{1}{c|}{-}                  & \multicolumn{1}{c|}{74.53}              & 623                          & \multicolumn{1}{c|}{-}                  & \multicolumn{1}{c|}{43216}              & 43530                        & \multicolumn{1}{c|}{0}                  & \multicolumn{1}{c|}{25}                 & 7                            \\ \hline
bf\_mmove\_1                                                                             & \multicolumn{1}{c|}{291.6}              & \multicolumn{1}{c|}{660.2}              & 64.63                        & \multicolumn{1}{c|}{0.025}              & \multicolumn{1}{c|}{0.005}              & 176                          & \multicolumn{1}{c|}{1}                  & \multicolumn{1}{c|}{2}                  & 11245                        & \multicolumn{1}{c|}{20006}              & \multicolumn{1}{c|}{16749}              & 28                           \\ \hline
bf\_mmove\_4                                                                             & \multicolumn{1}{c|}{245.3}              & \multicolumn{1}{c|}{578.2}              & 63.1                         & \multicolumn{1}{c|}{0.008}              & \multicolumn{1}{c|}{0.003}              & 159                          & \multicolumn{1}{c|}{1}                  & \multicolumn{1}{c|}{1}                  & 10070                        & \multicolumn{1}{c|}{20146}              & \multicolumn{1}{c|}{15165}              & 24                           \\ \hline
bf\_mmove\_7                                                                             & \multicolumn{1}{c|}{232}                & \multicolumn{1}{c|}{573}                & 66.31                        & \multicolumn{1}{c|}{0.007}              & \multicolumn{1}{c|}{0.005}              & 229                          & \multicolumn{1}{c|}{1}                  & \multicolumn{1}{c|}{3}                  & 15317                        & \multicolumn{1}{c|}{17010}              & \multicolumn{1}{c|}{14493}              & 15                           \\ \hline
bf\_mmove\_12                                                                            & \multicolumn{1}{c|}{257.3}              & \multicolumn{1}{c|}{508.2}              & 64.53                        & \multicolumn{1}{c|}{-}                  & \multicolumn{1}{c|}{182.14}             & 783                          & \multicolumn{1}{c|}{-}                  & \multicolumn{1}{c|}{92456}              & 50643                        & \multicolumn{1}{c|}{0}                  & \multicolumn{1}{c|}{15}                 & 6                            \\ \hline
oob\_1\_arr\_1                                                                           & \multicolumn{1}{c|}{278}                & \multicolumn{1}{c|}{598.8}              & 71.86                        & \multicolumn{1}{c|}{2.06}               & \multicolumn{1}{c|}{0.14}               & 55                           & \multicolumn{1}{c|}{556}                & \multicolumn{1}{c|}{83}                 & 3880                         & \multicolumn{1}{c|}{6121}               & \multicolumn{1}{c|}{6291}               & 39                           \\ \hline
oob\_1\_arr\_6                                                                           & \multicolumn{1}{c|}{308}                & \multicolumn{1}{c|}{591}                & 77.03                        & \multicolumn{1}{c|}{0.027}              & \multicolumn{1}{c|}{1.39}               & 103                          & \multicolumn{1}{c|}{14}                 & \multicolumn{1}{c|}{821}                & 8085                         & \multicolumn{1}{c|}{5541}               & \multicolumn{1}{c|}{6600}               & 28                           \\ \hline
oob\_1\_arr\_9                                                                           & \multicolumn{1}{c|}{284.6}              & \multicolumn{1}{c|}{571.2}              & 69.78                        & \multicolumn{1}{c|}{-}                  & \multicolumn{1}{c|}{273.8}              & 105                          & \multicolumn{1}{c|}{-}                  & \multicolumn{1}{c|}{155938}             & 7326                         & \multicolumn{1}{c|}{0}                  & \multicolumn{1}{c|}{12}                 & 27                           \\ \hline
oob\_1\_arr\_13                                                                          & \multicolumn{1}{c|}{297.2}              & \multicolumn{1}{c|}{507}                & 75.2                         & \multicolumn{1}{c|}{12.11}              & \multicolumn{1}{c|}{97.86}              & 207                          & \multicolumn{1}{c|}{3564}               & \multicolumn{1}{c|}{49165}              & 27241                        & \multicolumn{1}{c|}{254}                & \multicolumn{1}{c|}{686}                & 19                           \\ \hline
oob\_2\_arr\_1                                                                           & \multicolumn{1}{c|}{298.6}              & \multicolumn{1}{c|}{520.8}              & 73.53                        & \multicolumn{1}{c|}{-}                  & \multicolumn{1}{c|}{154.42}             & 117                          & \multicolumn{1}{c|}{-}                  & \multicolumn{1}{c|}{80080}              & 8558                         & \multicolumn{1}{c|}{0}                  & \multicolumn{1}{c|}{12}                 & 35                           \\ \hline
oob\_2\_arr\_5                                                                           & \multicolumn{1}{c|}{326.6}              & \multicolumn{1}{c|}{520.4}              & 71.1                         & \multicolumn{1}{c|}{-}                  & \multicolumn{1}{c|}{155.62}             & 165                          & \multicolumn{1}{c|}{-}                  & \multicolumn{1}{c|}{80662}              & 22759                        & \multicolumn{1}{c|}{0}                  & \multicolumn{1}{c|}{16}                 & 27                           \\ \hline
oob\_2\_arr\_8                                                                           & \multicolumn{1}{c|}{295.5}              & \multicolumn{1}{c|}{592.64}             & 69.8                         & \multicolumn{1}{c|}{-}                  & \multicolumn{1}{c|}{102.97}             & 188                          & \multicolumn{1}{c|}{-}                  & \multicolumn{1}{c|}{60384}              & 13366                        & \multicolumn{1}{c|}{0}                  & \multicolumn{1}{c|}{12}                 & 22                           \\ \hline
oob\_2\_arr\_13                                                                          & \multicolumn{1}{c|}{312.25}             & \multicolumn{1}{c|}{502.2}              & 70.95                        & \multicolumn{1}{c|}{-}                  & \multicolumn{1}{c|}{97.86}              & 192                          & \multicolumn{1}{c|}{-}                  & \multicolumn{1}{c|}{48694}              & 13401                        & \multicolumn{1}{c|}{0}                  & \multicolumn{1}{c|}{17}                 & 19                           \\ \hline
\textbf{Average}                                                                          & \multicolumn{1}{c|}{\textbf{283.43}}             & \multicolumn{1}{c|}{\textbf{567.53}}              & \textbf{68.34}                        & \multicolumn{1}{c|}{\textbf{2.8}}                  & \multicolumn{1}{c|}{\textbf{57.6}}              & \textbf{243}                          & \multicolumn{1}{c|}{\textbf{729.85}}                  & \multicolumn{1}{c|}{\textbf{30921}}              & \textbf{17481}                        & \multicolumn{1}{c|}{\textbf{6351}}                  & \multicolumn{1}{c|}{\textbf{5512}}                 & \textbf{22}                           \\ \hline
\end{tabular}

}

% \begin{tablenotes}
%     \centering
    %\scriptsize
   
% \end{tablenotes}
\vspace{-0.2cm}
\end{table*}

%So first, we utilize the IDE to login into the device, load the control application in the runtime, save it as a boot application and log out (Point 1). Creating a boot application enables {\FRAMEWORKNAME} to automatically restart the runtime after detecting a crash. Next, {\FRAMEWORKNAME} takes over the execution control while providing fuzzed input for each scan cycle. Since {\FRAMEWORKNAME} has control over the scan cycle, it does not drop any inputs due to a lack of synchronization with the scan cycle (Point 2). Finally, based on the error feedback received from the runtime, {\FRAMEWORKNAME} restarts the runtime and logs the crash input (Point 3).

% Fuzzing of control application binaries is a special case of component fuzzing since dedicated components control the execution of these binaries. The compiled control application runs in the thread spawned by the \texttt{SysTask} component, which is not exported to the network and thus cannot be influenced directly. Instead, {\FRAMEWORKNAME} fuzzes the binary inside the runtime context by controlling its execution through the \texttt{CmpApp} component. The latter offers complete control over start, stop, cold reset, and single-cycle operations with the runtime. 
% % Similarly, it also allows to reset the application in case of an exception in the control program that does not crash the runtime.
% Table~\ref{table:commands} shows the commands replicated for \texttt{CmpApp}.

We are now ready to demonstrate the application of FieldFuzz to the main goal of control application binary fuzzing. %Fuzzing control application binaries is a special case of runtime component fuzzing since dedicated components control the input delivery and execution to these binaries. 
The compiled control application runs in the thread spawned by the \texttt{SysTask} component, which is not exported to the network and thus cannot be influenced directly. Instead, FieldFuzz fuzzes the binary inside the runtime context by controlling its execution through the \texttt{CmpApp} component. The latter offers complete control over start, stop, cold reset, and single-cycle operations with the runtime. Table~\ref{table:commands} shows the commands replicated for \texttt{CmpApp}.

% For instance, \texttt{0x10}, \texttt{0x11} are used to start and stop the execution of the IEC binary.

% To set up the experiment, {\FRAMEWORKNAME} logs in to the device, and starts the control application. 
% % to the runtime, save it as a boot application and log out.
% % Creating a boot application enables {\FRAMEWORKNAME} to automatically restart the runtime after detecting a crash. 
% Next, {\FRAMEWORKNAME} takes over the execution control while providing fuzzing inputs for every scan cycle. Since {\FRAMEWORKNAME} has full control over the scan cycle, it does not drop any inputs due to a lack of synchronization with the scan cycle. Finally, based on the status feedback received from the runtime, {\FRAMEWORKNAME} logs the crash input.

To set up the experiment, FieldFuzz logs in to the device and starts the control application. Next, it takes over the execution control of the application while providing fuzzing inputs for every scan cycle. Since FieldFuzz has complete control over the scan cycle, it does not drop any inputs due to a lack of synchronization. Finally, it logs crash inputs based on the status feedback received.% from the runtime.

\noindent \textbf{Synthetic binaries.}
We use the same dataset of synthetic applications as used in ICSFuzz~\cite{proc:icsfuzz} for performing the experimental evaluation of FieldFuzz. The dataset comprises control applications written in Structured Text that include introduced vulnerabilities in their imported functions, such as buffer overflows and out-of-bounds write. These vulnerabilities exist due to missing bound checks in imported IEC 61131 library functions. Thus, the family of synthetic applications labeled in the dataset as \texttt{bf\_mmove} can cause a buffer overflow under certain conditions due to insufficient buffer size validation before calling a \texttt{SysMemMove} library function. Similar to the control application, this library is written in Structured Text. By looking deeper into its implementation in the runtime, we observe that \texttt{SysMem} component of the runtime provides the backend for this library and is implemented in C. The call of this wrapper, initiated by the control application, ends up in C code which triggers the native \texttt{memmove} function. For this reason, the crash in a vulnerable control application causes the failure of its thread and affects the entire runtime process (running with root privileges). Out-of-bounds write vulnerabilities involve an uninitialized array with a variable index manipulated to write at an arbitrary location. 
The numbers in the names of the vulnerable  binaries correlate with the complexity of the code. For instance, \texttt{bf\_mmove\_1} is the simplest initialization of the \texttt{SysMemMove}, while \texttt{bf\_mmove\_12} consists of multiple loops and conditional branching statements.

Table~\ref{table:icsfuzz_comparison} shows the results of fuzzing the control application binary and its comparison with ICSFuzz.
As the table demonstrates, on average FieldFuzz is $\approx$4.1x and $\approx$8.3x faster for arm32 and x64 (Intel) runtime variants, respectively, compared to ICSFuzz. The performance advantage of FieldFuzz comes from the communication protocol-based input delivery and complete control over the scan cycle. On the other hand, ICSFuzz incurs high latency and drops inputs during fuzzing when it misses the scan input cycle of the runtime. It should be noted that the number of crashes reported in Table \ref{table:icsfuzz_comparison} are a result of intentionally introduced vulnerabilities in the synthetic control applications used for performance evaluation. They concern solely the control applications, not the runtime itself.

% (a) How many runs have been performed in each fuzzing campaign, and what are the according averages and variances? 
% (b) When were the individual crashes detected in the fuzzing campaigns (Table 6 contains the time for the first crash only)? 
% (c) How are crashes grouped and deduplicated for FieldFuzz and ICSFuzz?
For the scope of this evaluation, we perform one run for each fuzzing campaign. Each campaign targets a distinct vulnerability in the control application binary. To provide metrics comparable to ICSFuzz, we do not perform additional grouping or deduplication of these crashes. We record the time of the first occurred crash in the campaign and increment the number of subsequent occurring crashes to produce the total metric within a one-hour time window.

It should also be emphasized that the measurements in Table~\ref{table:icsfuzz_comparison} are extracted for a single fuzzing instance for FieldFuzz. ICSFuzz requires a vendor-specific KBUS IO subsystem for input delivery, bounding itself to a physical device. Therefore ICSFuzz requires a physical device for fuzzing, which limits its scalability. On the other hand, FieldFuzz can parallelize fuzzing sessions by simply spawning multiple VMs.

Furthermore, FieldFuzz detects considerably more crashes than ICSFuzz, allowing it to cover a wider input space. On average, it detects $\approx$291x and $\approx$262x more crashes for the arm32 and x64 runtime variants, respectively, in the same one-hour fuzzing period. However, FieldFuzz detects fewer crashes for a select few samples across both variants. 
As mentioned previously, while higher-level components normally originate from exact same codebase, low-level system components can differ across various devices and architectures. We have observed that variants of memory management and exception handling implementation in device-specific system components can cause differences in crash behavior.
% As mentioned previously, low-level system components are implemented differently across various devices and architectures, resulting in different bugs and vulnerabilities. 
For example, in our 32 bit runtime variant, we observe that the \texttt{SysMem} component prevented the runtime from crashing for some samples and instead wrote ``Operation not permitted" in the logs, successfully sanitizing the input.

\section{Discussion and Limitations}
%\noindent \textbf{Security mitigations by the runtime.} Codesys implements a set of security controls to mitigate attacks on the runtime. For instance, the latest runtime version enables the User Management feature by default, thwarting unauthorized login into the PLC. 

% First, the fuzzing approach is compared to ICSFuzz (results shown in Table 6). It appears that for this experiment, the target platforms between ICSFuzz (target: Wago PFC100) and FieldFuzz (targets: BeagleBone and Linux x64) differ. This greatly impacts the validity of this experiment and contests the conclusions drawn from it. Why is it not possible to run the proposed framework also on the Wago device?

% Third, I am missing an evaluation of the induced overhead. In my opinion, the Ghost component can take up significant memory and compute resources on the devices. This could clash with the memory and real-time constraints of the device and make the application unfit for use. I would appreciate if the authors could discuss and, if possible, evaluate this aspect.
\noindent \textbf{Evaluation limitations: }
During the evaluation, we compared FieldFuzz to ICSFuzz since it is the only state-of-the-art work specifically targeting ICS devices. However, While ICSFuzz was designed to run on the device itself, FieldFuzz interacts with remote devices over the network. Due to this architectural difference, it is not possible to compare the performance on the same hardware platform from the computational efficiency point of view. 
Therefore, in this evaluation, we aim to compare these two systems in their ability to discover vulnerabilities in the control programs and the runtime by common metrics. Moreover, as the Ghost coverage mechanism injects logic and operates on the fuzzed device itself, it can potentially consume significant memory and compute resources of these constrained devices. This adds a potential risk to meeting the real-time constraints of the PLC operation when fuzzing custom component functions with a larger codebase. A more thorough performance evaluation of this aspect is required and is a direction for future work. Next, the scope of our evaluation included a single run per fuzzing campaign (runtime component or control application binary) and did not address the potential run-to-run performance variety of the fuzzer. Finally, in the chosen metrics and the evaluation approach, we have mainly targeted to produce the metrics that are compatible with ICSFuzz, as was discussed in Section~\ref{sec:results}. %\cite{DBLP:journals/corr/abs-1808-09700} \hl{.}

\noindent \textbf{Runtime security mitigations.} The latest runtime version enables the User Management feature by default, thwarting unauthorized login into the PLC. However, out-of-the-box credentials are the default unless manually changed, while the communication is not encrypted. The runtime also expects the client to perform the \texttt{Login} action with \texttt{CmpDevice} for establishing a session, but this process does not involve actual authentication. Furthermore, the security mitigation properties of the runtime executable differ among platforms. For instance, the WAGO PFC200 PLC (with firmware 03.00.39(12)) used in our setup contains the runtime that is compiled without all of the typical exploit mitigations (no Relocation Read-Only (RELRO), stack canary, NX bit, or Position Independent Executable (PIE)). Finally, the monitoring bytecode interpreter, involved with input delivery to the control application by performing extensive memory operations, applies its own memory access checks. For each execution, before loading the bytecode, the interpreter sets a canary to ensure the integrity of the stack. However, we found that this canary has a fixed value of \texttt{0x5AF096A5} regardless of the target platform, which defeats its purpose.

\noindent \textbf{Runtime component coverage.} Maximizing component coverage through different mutation strategies is orthogonal to our work and can be explored as a future research direction. Methodologies such as SNOOZE~\cite{proc:banks2006snooze}, program-adaptive mutational fuzzing \cite{proc:program_adaptive}, and PAVFuzz \cite{proc:pavfuzz} can be integrated with FieldFuzz to improve coverage.

\noindent \textbf{Black-box fuzzing challenges.} FieldFuzz does not require access or any modifications to the controller, ensuring the universality and scalability of the proposed approach. However, this incurs limited code coverage information. We rely on the retrieved status codes to partially address this for runtime components for understanding the execution path. We have found that the debugging capabilities of the full-featured VM can emulate the functionality of the service layer without requiring actual network transmission. This requires pre-loading a harness as a shared library into the runtime and hooking the authentication and packet processing functions in the runtime process. This approach builds a more traditional and comprehensive fuzzing approach combined with full-featured code coverage. However, in the context of ICS, such a white-box fuzzing approach has substantial limitations:

\begin{enumerate}
[leftmargin=*, topsep=0pt,itemsep=-1ex,partopsep=1ex,parsep=1ex]
\item The compiled harness and fuzzing instance is tied to one specific target platform (architecture), while some vulnerabilities are platform-specific, reducing generalization.
\item This approach is possible with a SoftPLC build of the runtime on top of a typical desktop-grade VM. However, real-world COTS devices hardly have such extensive debugging and instrumentation capabilities. 
\item Full shell access for controlling the device is rare, as many ICS devices embed the runtime on legacy RTOS or use bare-metal runtime variations. 
Gaining white-box fuzzing capabilities would require re-flashing the controller with a modified kernel image and relying on remote debugging. 
\end{enumerate}

\section{Conclusion}

This paper presents FieldFuzz -- a fuzzing framework for control applications and industrial runtimes, capable of discovering vulnerabilities in over 400 known ICS devices from 80 industrial device vendors. It facilitates efficient network-based fuzzing by i) reverse-engineering enabled remote control of control applications and runtime components, 
ii) automated command discovery and status code extraction via network traffic and iii)  a monitoring setup to allow on-system tracing and coverage computation. It is the first fuzzer to be able to fuzz control applications in situ, within their execution context of the runtime, even on target blackbox devices. 
We successfully fuzz the various runtime instances (on different architectures and by different vendors) of the Codesys runtime, reporting three CVEs. In addition, FieldFuzz achieves a speedup of $\approx$8.3x compared to the state-of-the-art for control application binaries and an increased crash discovery of $\approx$291x and $\approx$262x for 32 and 64 bit runtime variants, respectively. We perform fuzzing on physical and virtualized ICS devices to demonstrate automation capabilities, reliability, and performance improvements against the current state-of-the-art. With FieldFuzz, we provide researchers with a robust open-source framework to enable future research in this direction.

\section*{Reported CVEs \& Published Tools}
\input As a result of this work, our reported vulnerabilities were assigned \texttt{CVE-2022-22514}, \texttt{CVE-2022-22508}, and \texttt{CVE-2022-22507}. We release \href{https://github.com/fridgebuyer/FieldFuzz}{FieldFuzz}~\cite{repo:FieldFuzz}, \href{https://github.com/fridgebuyer/FieldFuzz-Ghost}{Ghost}~\cite{repo:FieldFuzz-Ghost}, and the \href{https://github.com/fridgebuyer/codesys3-dissector}{Wireshark dissector for Codesys v3 protocol}~\cite{repo:codesys3-dissector} as open source tools with this work.    

\begin{acks}
Constantine Doumanidis has been supported by the NYU Abu Dhabi Center for Cyber Security, and Prashant Rajput has been supported by the NYU Abu Dhabi Global PhD Fellowship.

Jianying Zhou's research is supported by the National Research Foundation, Singapore, under its National Satellite of Excellence Programme “Design Science and Technology for Secure Critical Infrastructure: Phase II”. Any opinions, findings and conclusions or recommendations expressed in this material are those of the author(s) and do not reflect the views of National Research Foundation, Singapore.
\end{acks}

% \section*{Reported CVE IDs}
% As a result of this work, our reported vulnerablities were assigned \texttt{CVE-2021-34604}, \texttt{CVE-2022-22508}, and \texttt{CVE-2022-22507}.

% {\footnotesize \bibliographystyle{ieee}
% \bibliography{z-ics}}

\bibliographystyle{ACM-Reference-Format}
\bibliography{z-ics}

%\printbibliography

\appendix
% \input{Tables/commands}
% \subsection{Runtime Component Structure}\label{appendix:cmpstructure}
% Each Codesys component follows a specific programming pattern in order to be integrated with the runtime. This structure is as follows:
% \begin{itemize}[leftmargin=*, topsep=0pt,itemsep=-1ex,partopsep=1ex,parsep=1ex]
%     \item It contains an entry function with a standard C struct passed as the argument. One example of such entry function for \texttt{CmpMonitor2} component is shown in Listing~\ref{code:cmpmonitor2_entry}, which we renamed as \texttt{CmpMonitor2\_\_Entry}.
%     \item Declares its internal ID (in this case 50) and passes a string literal of its name to the Component Manager.
%     \item Calls standardized import and export functions that enable inter-component communication (Lines 4-5).
%     \item Handles standardized create, delete event hooks and a version identifier function (Lines 6, 8, and 9).
%     \item Subscribes for custom events sent by other components using a pointer to an event handler function (Line 7).
% \end{itemize}

% \input{Code/cmpmonitor2_entry}

% \renewcommand\thesection{\Alph{section}}

% \newpage

% \section{Status Codes}
% \input{Tables/errorcodes}

%\input{Text/CodesysStack}

\end{document}